\documentclass[11pt]{article}
\usepackage{amssymb,amsfonts,amsmath,amsthm,mathrsfs}
 \usepackage[english]{babel}
 \usepackage[utf8]{inputenc}  
\usepackage{enumitem}
 \usepackage{url}
 \usepackage{graphicx}
 \usepackage{color}
 \usepackage{empheq,tikz}
 \usepackage{vmargin}
 \usepackage{minitoc}
 \usepackage[a4paper, margin=1in]{geometry}
 \usepackage{pdfpages} 
\usepackage{afterpage}
\usepackage{bm}
\usepackage{dsfont}
\usepackage{hyperref}
\usepackage{subcaption}
\usepackage{xcolor}
\usepackage{authblk}

\usepackage{cleveref}

\usepackage[most]{tcolorbox} 

\newtcolorbox[auto counter]{optionalnote}[2][]{
    parbox=false,
    colbacktitle= white,
    colback=green!5!white,
    colframe=white!45!black,
    coltitle=black,
    enhanced,
    attach boxed title to top left={yshift=-1mm},
    title={\thetcbcounter.~#2}
,#1}

\newtcolorbox{highlight-result}[1][]{
 parbox=false,
 boxrule=0pt,top=0pt,bottom=0pt,
colback=blue!7!white,
enhanced,#1}

\tcbset{colback=blue!7!white, colframe=white!50!white,top=0mm,bottom=0mm,right=0mm,left=0mm}

\usepackage[
backend=biber,
style=alphabetic,
sorting=ynt
]{biblatex}

\usepackage[textsize=normalsize,textwidth=2cm,colorinlistoftodos]{todonotes}
\newcounter{todocounter}

\usepackage{fancyhdr}

 \AtEndDocument{\label{lastpage}}
\setmargrb
 {2.5cm} 
 {1.5cm} 
 {2.5cm} 
 {2cm} 
 
\newcommand{\ensem}[1]{\langle #1 \rangle}

\title{Acoustic waves in a halfspace material filled with random particulate}
\author{Paulo S. Piva, Kevish K. Napal, Art L. Gower}
\affil{Department of Mechanical Engineering, The University of Sheffield, UK}
\date{\today}

\begin{document}

\tableofcontents

\maketitle

\begin{abstract}
Particulate materials include powders, emulsions, composites, and many others. This is why measuring these has become important for both industry and scientific applications. For industrial applications, the greatest need is to measure dense particulates, in-situ, and non-destructively. In theory, this could be achieved with acoustics: the standard method is to send an acoustic wave through the particulate and then attempt to measure the effective wave speed and attenuation. A major obstacle here is that it is not clear how to relate the effective wave speed and attenuation to the reflection and transmission coefficients, which are far easier to measure. This is because it has been very difficult to mathematically account for different background mediums. In this paper we resolve this obstacle. We present how to account for different background mediums for a simple case, to help comprehension:   
    a halfspace filled with a random particulate, where the background of the halfspace is different to the exterior medium. 
    The key to solving this problem was to derive a systematic extension of a widely used closure approximation: the quasi-crystalline approximation (QCA).
    We present some numerical results to demonstrate that the reflection coefficient can be easily calculated for a broad range of frequencies and particle properties.
\end{abstract}

\section{Introduction}
\label{sec:intro}


Particulate materials are composed of small particles embedded in a continuous medium, like a powder in air, emulsions such as oil droplets in water, or air bubbles appearing in boiling water.
Many applications, for example sensing, require a descriptive model of particulate materials. The positions of the particles are disordered and are either not known or cannot be completely controlled. Therefore, the most important features of particulate materials are statistical, such as the average particle size or inter-particle distance.


\textbf{Sensing.} Non-destructive measurements of particle size and properties can be achieved using wave scattering (acoustic, electromagnetic or elastic) \cite{ISO_2017,Challis_2005,Al-Lashi_2015,FORRESTER_2016}. Characterising, or monitoring, the particles is needed to 
ensure quality, or can be used in feedback loops during production or manufacturing. 
As the particles can change position, and even properties, in time (or space), 
to obtain a reliable measurement experiments need to be repeated many times to then compute the average wave response. This average response depends only on the statistical properties, such as average particle size, which are usually the most important features for industrial applications.  Mathematically, the result of averaging over measurements in time (or space) can be equivalent to a procedure called ensemble averaging, which we use in this work, see \cite{Foldy_1945,MISHCHENKO_2016, Huang_1963} for details.



\textbf{Broad frequency range.} To sense the size of particles, we need to consider a broad range of frequencies, so that the wavelengths are comparable to the particle size. It is not enough to develop a theory for only the low frequency limit, as in this limit it is not possible to sense particle size distribution. For example, in the long wavelength limit (low frequency) for acoustics \cite{Caleap_2012}, the material is completely described by two numbers: the effective density and effective bulk modulus \cite{gower_reflection_2018}. So, in terms of sensing, to learn more than just two numbers from a wave experiment we need to consider shorter wavelengths.


\textbf{Effective waves method.} As we need a broad range of frequencies, we make use of a method called the \emph{Effective waves method} \cite{gower2021effective,Gower_2023,Kevish2024}, which can account for wave scattering in dense or sparse particulate materials for a broad range of frequencies (see \cite{Aris2024} for phase diagrams). This method accounts for multiple scattering between all particles, 
and provides a way to perform sensitivity studies on a wide range of parameters such as particle size distribution and volume fraction. 

\begin{figure}[h!]
    \centering
    \includegraphics[width=0.9\textwidth]{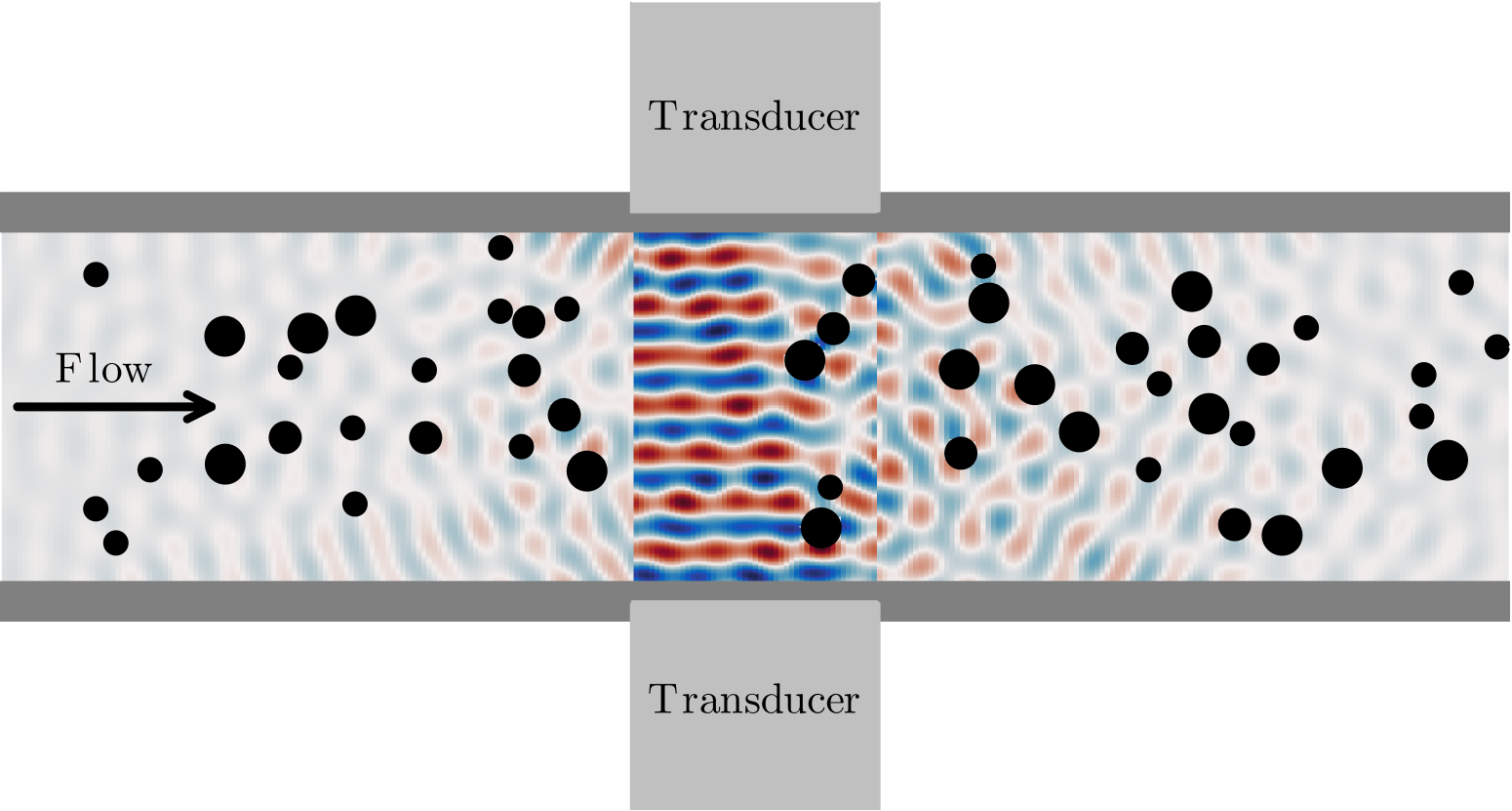}
    \caption{Shows the cross-section of a pipe, with a fluid flowing in the direction of the black arrow. Particles are suspended in the fluid, represented by black circles. A pair of transducers (acoustic sensors) are attached to the pipe walls. One transducer emits waves and measures their reflection, while the other measures the transmitted wave. This illustration is just a pictorial representation. For any real applications, particles would be much smaller and more numerous. The figure was generated in Julia with the library \cite{2020MultipleScatering.jl}.}
    \label{fig:pipe}
\end{figure}

\textbf{Different background mediums.} When using the Effective waves method to design ways to sense particles we came across a significant barrier: it is not clear how to calculate the average scattering when the source comes from a medium which is different from the background medium of the particles. An example is shown in \Cref{fig:pipe}, which depicts an emulsion (particles in a fluid) travelling along a metal pipe.
The pipe material is different from the background fluid. There have been work, and experiments, in the literature that considers the case of different background mediums \cite{fawcett2021effective,Garcia-Valenzuela:05,Simon_tony2024}, but the expressions used do not come from first principles calculations. We discuss this further in the literature review below. 
To design robust sensing methods, we need to account for these different mediums, which is the main goal of this paper. After significant calculations from first principles, we arrive at a simple strategy which will lead to more robust sensing methods.

\textbf{Quasi-crystalline approximation.} To solve for all orders of multiple scattering between particles, on average, one needs to use a closure approximation \cite{Closure1971,2016_closure}. The most standard closure assumption to account for scattering between particles is called the Quasi-Crystalline Approximation (QCA) \cite{Lax_QCA_1952, Varadans_84,gower2019multiple}. We derive a consistent extension to QCA (named X-QCA) to also account for scattering between particles and interfaces. To summarise, X-QCA accounts for the same scattering orders as QCA and leads to simpler calculations.
For clarity, we consider only a simple case: plane wave incidence on a halfspace filled with random particles (see \Cref{fig:multiple_scattering}), and only for acoustic wave scattering.

\textbf{Further applications.} Other than sensing, accounting for scattering between layers and particles can lead to improved design of: graded particulate materials \cite{Miyamoto_1999} and disordered metamaterials with tailored frequency response. Disordered particulate materials can be far simpler to manufacture on a large scale, because the exact positioning of the particles does not need to be carefully controlled, as it does in most periodic metamaterials.

\textbf{A brief literature review.} Most of the work done in particulates and composites is focused on the low frequency limit, for on example \cite{Parnell_2010}, or for broader frequencies but with only one background medium \cite{Caleap_2012,Caleap_2015,Linton_2005,Linton_2006}.


Recent work on acoustic scattering by random composite media has been carried out by John R. Willis \cite{Willis2019, Willis2020, Willis2023}, in which a broad frequency response is considered. According to \cite{Willis2019}, the 
model framework is based on elasticity \cite{Willis1981}, and it describes wave scattering (acoustic or elastic) from a halfspace formed by three distinct phases. Compared to \Cref{fig:setting-halfspace} later in the text, one phase would be represented by the exterior medium (blue), the second could be thought of as the matrix (yellow), and the last as the particles (black circles). However, the distribution of each phase in the halfspace is given in terms of a two-point correlation function, and not by placing particles as shown in \Cref{fig:pipe,fig:setting-halfspace}. To solve the resulting equations, Willis assumed two of the phases have the same bulk modulus in \cite{Willis2023}. In this paper, 
we do not impose any restrictions on the acoustic properties of the three phases.

In this work we follow a first principles approach similar to \cite{Slab_book}, which accounts for all orders of multiple scattering. 
To solve for the average reflected and transmitted waves, \cite{Slab_book} needed to use the standard QCA (for scattering between particles) and specialise to either a low particle volume fraction or low frequency. In contrast, in this paper, we do not need to specialise to a low particle volume fraction or low frequency and reach solutions which are easier to compute, and are, in principle, as accurate. To achieve this we deduce an extension of the quasi-crystalline approximation (X-QCA), which simplifies the scattering between particles and walls.


\textbf{Summary of the paper.} In Section \ref{subsec:strat} we introduce an overview of how to account for multiple scattering between particles and interfaces for the average wave in materials with random microstructure. In Section \ref{sec:setting} we define the setup of a plane wave incident on a halfspace filled with particles. In Section \ref{sec:one_configuration} we explicitly write the system of equations of the problem for acoustic scattering of waves for one configuration of particles. In Section \ref{sec:ensemble} we define the probability density of each realisation, and all the statistical assumptions used throughout the paper. In Section \ref{sec:ensem_avrg_fields} we compute the average of the total pressure field and apply boundary conditions on the interface between the different mediums in the halfspace. In Section \ref{sec:backscattering-operator} we use X-QCA to determine the average of the backscattering operator to make a clear connection with the strategy introduced in Section \ref{subsec:strat}. In Section \ref{sec:qca} we derive X-QCA, which can be used in the presence of different background mediums. In Section \ref{sec:effective_waves} we apply the Effective waves method and present the numerical results achieved.

\subsection{Overview of the strategy}
\label{subsec:strat}

In this section, we show how to intuitively deduce wave scattering from a random particulate in the presence of different background mediums. We do this for the simplest scenario: plane wave scattering. After this section we deduce rigorously the results presented here.

Let $\bm{r} = (x, y, z) \in \mathbb R^3$ be a position. Consider a homogeneous acoustic halfspace, which we call the background matrix, occupying the the region $z>0$, in $\mathbb{R}^3$, which is filled with a random complex material. 

Consider another homogeneous acoustic halfspace $z<0$ which has different properties to the background matrix in $z>0$. From the region $z<0$, an incident plane wave propagates in the positive $z$ direction given by
\begin{equation}
    \notag
    u_{\text{in}} (\bm{r}) = G \mathrm e^{\mathrm i k z}
\end{equation}
with $k > 0$ being the wavenumber of the $z<0$ region.

\begin{figure}[ht]
\centering

\begin{subfigure}[b]{.55\textwidth}
\centering

\begin{tikzpicture}

\draw[color = blue!10, fill=blue!10]
(0,0) -- (-4,0) -- (-4,3) -- (0,3) -- cycle;

\draw[color = yellow!40, fill=yellow!40]
(0,0) -- (5,0) -- (5,3) -- (0,3) -- cycle;

\fill[color = yellow!40] (3,0) -- (5,0) -- (5,3) -- (3,3) -- cycle;

\draw[-] (0,3) -- (0,0);
\draw[-, dashed] (3,3) -- (3,0);

\filldraw[color=black] (3.5,2.7) circle (2pt);
\filldraw[color=black] (4.1,1.8) circle (3pt);
\filldraw[color=black] (4.8,2.4) circle (2.7pt);
\filldraw[color=black] (3.3,0.3) circle (2pt);
\filldraw[color=black] (3.8,1.1) circle (2.6pt);
\filldraw[color=black] (4.7,0.7) circle (2.8pt);

\draw[->, blue, thick] (-3,1.9)--(-1,1.9);
\node at (-2,2.2) {\textcolor{blue}{$G$}};

\draw[->, magenta, thick] (-1,1.1)--(-3,1.1);
\node at (-2,1.4) {\textcolor{magenta}{$\langle R \rangle$}};

\draw[->, red, thick] (2.5,1.1)--(0.5,1.1);
\node at (1.5,1.4) 
 {\textcolor{red}{$\langle B \rangle$}};

\draw[->, green!60!black, thick] (0.5,1.9)--(2.5,1.9);
\node at (1.5,2.2) 
 {\textcolor{green!60!black}{$\langle A \rangle$}};

\draw[<->, thick] (0,0.15)--(3,0.15);
\node at (1.5,0.4) {$\delta$};

\end{tikzpicture}

\caption{Case $\delta > 0$}
\end{subfigure}
\hfill
\begin{subfigure}[b]{.43\textwidth}
\centering

\begin{tikzpicture}

\draw[color = blue!10, fill=blue!10]
(0,0) -- (-4,0) -- (-4,3) -- (0,3) -- cycle;

\fill[color = yellow!40] (0,0) -- (2,0) -- (2,3) -- (0,3) -- cycle;

\draw[-] (0,3) -- (0,0);

\filldraw[color=black] (0.5,2.7) circle (2pt);
\filldraw[color=black] (1.1,1.8) circle (3pt);
\filldraw[color=black] (1.8,2.4) circle (2.7pt);
\filldraw[color=black] (0.3,0.3) circle (2pt);
\filldraw[color=black] (0.8,1.1) circle (2.6pt);
\filldraw[color=black] (1.7,0.7) circle (2.8pt);

\draw[->, blue, thick] (-3,1.9)--(-1,1.9);
\node at (-2,2.2) {\textcolor{blue}{$G$}};

\draw[->, magenta, thick] (-1,1.1)--(-3,1.1);
\node at (-2,1.4) {\textcolor{magenta}{$\langle R \rangle$}};

\end{tikzpicture}

\caption{Case $\delta \to 0$}
\end{subfigure}

\caption{An illustration of the amplitudes of different plane waves. The blue on the left is the homogeneous halfspace from where the source ($G$) originates, and the reflected wave ($\ensem{R}$) heads into. The region on the right of both (a)  and (b) with the particles is a homogeneous matrix with an embedded random complex material (shown as particles here). The middle of (a) is a yellow layer with thickness $\delta$, which is homogeneous, and has the same properties as the background matrix of the halfspace filled with particles.}
\label{fig:halfspace}
\end{figure}
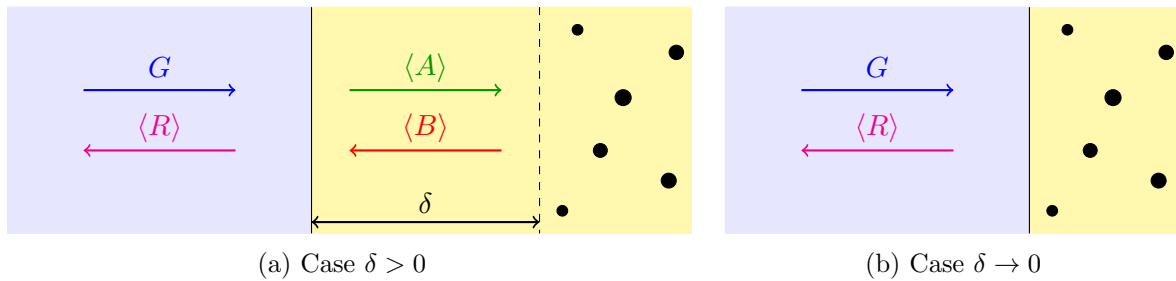

Surprisingly, to describe the average transmitted and reflected waves due to the incident plane wave is not straightforward, and arguably unsolved when considering all the multiple scattering between the interface and embedded random medium. 
During our work, we realised a simple intuitive trick to arrive at the same results achieved by the first principles calculations. The idea is to use what is already known: the solution of an average reflected plane wave from a complex material that is embedded in just one homogeneous medium. To use this solution, we consider an artificial region $0 < z < \delta$ which is homogeneous and has the same properties as the background matrix. See \Cref{fig:halfspace}a for an illustration.


For just one realisation of the complex material, i.e. the deterministic case, the total acoustic field is given by
\begin{equation}
    \label{eq:total_wave}
    u_{\text{tot}} (\bm{r}) = 
    \left\{
    \begin{aligned}
        G \mathrm e^{\mathrm i k z} + R \mathrm e^{-\mathrm i k z} + \varepsilon_-(\bm{r}), \quad & z<0
\\
        A \mathrm e^{\mathrm i k_0 z} + B \mathrm e^{-\mathrm i k_0 z} + \varepsilon_+(\bm{r}), \quad & 0<z<\delta
    \end{aligned}
    \right.
\end{equation}
where $k_0$ is the wavenumber of the region $0<z<\delta$, and $G, R, A, B \in \mathbb C$ are the amplitudes of the: incident plane wave, reflect plane wave, and two transmitted plane waves, respectively. The terms $\varepsilon_{\pm} (\bm{r})$ represents the non-planar contribution from the random material. This non-planar contribution will be zero later when taking an ensemble average over the random variables.  

We can calculate the unknown amplitudes by applying standard transmission boundary conditions across $z = 0$
given by %
\begin{equation}
\label{eq:trans_bound_conditions}
    \left\{
    \begin{aligned}
        &u_{\text{tot}} (\bm{r}) \text{ is continuous at } z = 0,
\\
        &\frac{1}{\rho(\bm{r})} \frac{\partial}{\partial z}  u_{\text{tot}} (\bm{r}) \text{ is continuous at } z = 0,
    \end{aligned}
    \right.
\end{equation}
where $\rho(\bm{r})$ is the density of the medium. 


Let us attempt to use the above to deduce the average reflection. To describe averages we introduce the random variable $\sigma$ to denote one realisation (or configuration) of this random material. For example, in a material composed of small particles $\sigma$ represents one possible configuration of the particles and their acoustic properties, see \Cref{fig:halfspace}b. In this sense, the ensemble average $\ensem{\circ}$ gives the average of $\circ$ over all possible realisations $\sigma$. Later we define this in detail.

Returning to \eqref{eq:total_wave}, we know that on average $\langle \varepsilon_{\pm} (\bm{r}) \rangle = 0$ due to planar symmetry. Performing the ensemble average of both sides of \eqref{eq:total_wave} then leads to
\begin{equation}
    \notag
    \langle u_{\text{tot}} (\bm{r}) \rangle = 
    \left\{
    \begin{aligned}
        G \mathrm e^{\mathrm i k z} + \langle R \rangle \mathrm e^{-\mathrm i k z}, \quad & z<0,
\\
        \langle A \rangle \mathrm e^{\mathrm i k_0 z} + \langle B \rangle \mathrm e^{-\mathrm i k_0 z}, \quad & 0 < z < \delta,
    \end{aligned}
    \right.
\end{equation}
where $\langle G \rangle = G$ because the incident wave is the same for each realisation $\sigma$. \Cref{fig:halfspace}a shows how each plane wave contributes to the average field. The goal is to first solve the case shown in \Cref{fig:halfspace}a and then take the limit $\delta \to 0$ to reach the solution of the case shown in \Cref{fig:halfspace}b.

%

Taking an ensemble average on both side of the boundary conditions \eqref{eq:trans_bound_conditions}, after some algebra, results in
\begin{equation}
    \label{eq:system_avrg}
    \left\{
    \begin{aligned}
        & \langle R \rangle = \zeta_R \, G + \zeta_T \, \langle B \rangle,
\\
        & \langle A \rangle = \gamma_0 \zeta_T \, G - \zeta_R \, \langle B \rangle,
    \end{aligned}
    \right.
\end{equation}
with $\zeta_R$, $\zeta_T$ and $\gamma_0$ being constants that depend on the material properties of the background mediums and are provided in Section \ref{subsec:boundary_conditions_hs}.

With the above, we have 3 unknowns: $\ensem{R}$, $\ensem{B}$, $\ensem{A}$, but only 2 equations. We need another equation. We can obtain another equation by knowing how the particles themselves reflect a plane wave. The transmitted wave with amplitude $A$ is reflected, in some sense, by the particulate material in the region $z> \delta$, and creates the reflected wave with amplitude $B$. This reflection is linear and can be represented by some scalar $\mathbb T_{\sigma}$ such that
\begin{equation}
    \label{def:bs_operator}
    \mathbb T_{\sigma} \, A := B.
\end{equation}
We call $\mathbb T_{\sigma}$ the backscattering operator, and it depends on each realisation $\sigma$.
See \cite{gower2019multiple,gower2021effective} for examples of the average of this operator. 

The main issue now is that taking an ensemble average on both side of \eqref{def:bs_operator} results in
\begin{equation}
    \label{eq:system_avrg_2}
        \langle B \rangle = \langle \mathbb T_{\sigma} \, A  \rangle,
\end{equation}
which does provide another equation, but also delivers another unknown $\langle \mathbb T_{\sigma} \, A  \rangle$, which can not be written directly in terms of $\langle A  \rangle$. This is because the waves $A$ and $B$ have been reflected between the complex random material and the interface at $z=0$, so both of these waves do depend on the realisation $\sigma$.
If the background matrix medium (yellow) was the same as the exterior (blue) medium, both shown in \Cref{fig:halfspace}, then $A$ would be the incident wave ($A=G$) and we would have $\ensem{\mathbb T_{\sigma} \, A} = \ensem{\mathbb T_{\sigma} } A$, as the incident wave does not depend on the realisation $\sigma$. 

To resolve this, it is normal to assume a closure relation \cite{Closure1971,2016_closure}.
The simplest and most commonly used is a naive mean field approximation \cite{kolomietz2020mean}, given by
\begin{equation}
    \label{eq:approx_interface}
    \langle B \rangle = \langle \mathbb T_{\sigma} \, A  \rangle \approx \langle \mathbb T_{\sigma} \rangle \langle A  \rangle.
\end{equation}
At first, this approximation may appear crude. However, we will show in this paper that the above approximation equivalent to the Quasi-crystalline approximation (QCA).
Therefore, it would not be useful, or consistent, to use a more accurate approximation when already QCA itself is assumed.  

The main goal of this work is to show that closure approximations of the form \eqref{eq:approx_interface} can be deduced from first principles when using the same assumptions as QCA.
Beyond just scattering between halfspaces, our approach leads to a general strategy to calculate (on average) multiple scattering between random particles and different background mediums.



\section{Setting of the problem}
\label{sec:setting}

Our aim is to describe wave scattering from a halfspace,
\begin{equation}
    \notag
    \mathcal R := \left\{ \bm r = (x,y,z) \in \mathbb R^3 \, | \, z \geq 0 \right\},
\end{equation}
filled with particles.  The region that all particles occupy is denoted by $\mathcal P \subset \mathcal R$, which is the union of non-overlapping homogeneous spheres with radius $a > 0$, sound speed $c_s \in \mathbb{C}$ and density $\varrho_s \in \mathbb{R}$\footnote{It is not difficult to generalize all results presented here for the multispecies case, where each particle can have a different radius, sound speed and density. The procedure would be the same as done in \cite{gower2021effective}.}.

\begin{figure}[h!]
\centering
\begin{tikzpicture}
\draw[color = blue!10, fill=blue!10]
(-12,-2.2) -- (-1.8,-2.2) -- (-1.8,2.2) -- (-12,2.2) -- cycle;
\draw[color = yellow!40, fill=yellow!40]
(-1.8,-2.2) -- (2,-2.2) -- (2,2.2) -- (-1.8,2.2) -- cycle;
\draw[-] (-1.8,-2.2) -- (-1.8,2.2);
\coordinate (A) at (0,.5);
\coordinate (B) at (-1.2,0.25);
\coordinate (C) at (1,1);
\coordinate (D) at (-0.75,1.25);
\coordinate (E) at (-0.5,-0.7);
\coordinate (F) at (0.9,-1.2);

\filldraw (A) circle (2pt);
\filldraw (B) circle (3pt);
\filldraw (C) circle (2.7pt);
\filldraw (D) circle (2pt);
\filldraw (E) circle (2.6pt);
\filldraw (F) circle (2.8pt);
\draw (0,-2) node[above]{$\mathcal{R}$};
\draw (A) node[below]{1};
\draw (B) node[below]{2};
\draw (C) node[below]{3};
\draw (D) node[above]{4};
\draw (E) node[below]{5};
\draw (F) node[below]{6};

\node[draw,text width=4cm,text centered,minimum width=3cm,minimum height=1.2cm, fill=white] at (-9.5,0) {Exterior medium:\\sound speed $c$\\ density $\varrho$};
\draw[<-] (-1.2,-1)--(-3,-1);
\node[draw,text width=4cm,text centered,minimum width=3cm,minimum height=1.2cm, fill = white] at (-4.5,-1) {Matrix: $c_0,\varrho_0$};
\draw[<-] (0.88,1)--(-3,1);next
\node[draw,text width=4cm,text centered,minimum width=3cm,minimum height=1.2cm, fill = white] at (-4.5,1) {$i$-th particle: $c_s,\varrho_s$\\radius $a$};
\end{tikzpicture}
\caption{Cross-section of a homogeneous halfspace $\mathcal R$, filled with homogeneous spherical particles. The set of all points in particles is denoted by $\mathcal P$. The acoustic properties of each medium (sound speed and density) are specified, together with particle radius.}
\label{fig:setting-halfspace}
\end{figure}
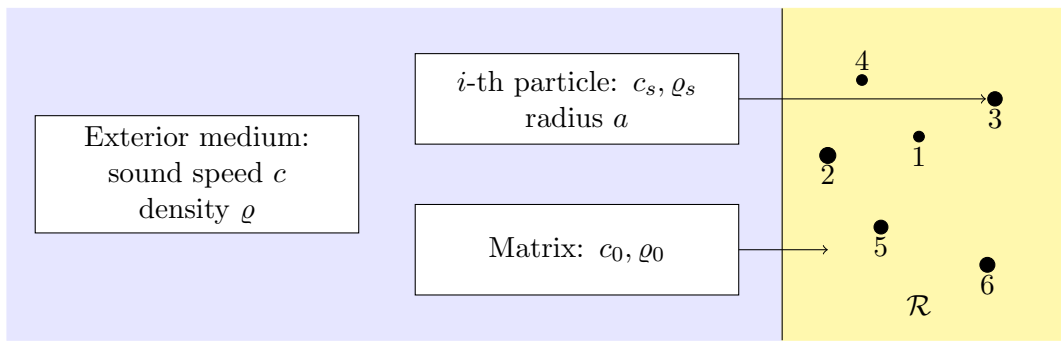

We call the region in between all the particles, $\mathcal R \setminus \mathcal P$, the matrix, and the region $z < 0$ the exterior medium, from where the incident wave originates. See \Cref{fig:setting-halfspace} for an illustration.
The exterior medium has a speed of sound $c \in \mathbb{R}$ and density $\varrho \in \mathbb{R}$, while the matrix has homogeneous acoustic properties, $c_0 \in \mathbb{R}$ and $\varrho_0 \in \mathbb{R}$.

%


The pressure field in the frequency domain is denoted by $u_{\text{tot}}$ and it satisfies the following Helmholtz equations:
\begin{equation}
    \notag
    \begin{aligned}
        \nabla^2 u_{\text{tot}}(\bm{r}) + k^2 u_{\text{tot}}(\bm{r}) = 0,& \quad \text{for } \;\; \bm{r} \notin \mathcal{R},
\\
        \nabla^2 u_{\text{tot}}(\bm{r}) + k_0^2 u_{\text{tot}}(\bm{r}) = 0,& \quad \text{for } \;\; \bm{r} \in \mathcal R  \setminus \mathcal P,
    \end{aligned}
\end{equation}
where $\omega \in \mathbb R$ is the angular frequency, $k := \omega / c$ is the wavenumber for the exterior medium (blue region), $k_0 := \omega / c_0$ is the wavenumber for the matrix (yellow region). 

The total field in the exterior region equals the incident wave plus a reflected field, 
\begin{equation} \label{eq:total_field_medium}
    u_{\text{tot}}(\bm{r}) = u_{\text{in}}(\bm{r}) + u_{\text{rf}}(\bm{r}) =
        G \mathrm e^{\mathrm i \bm{k} \cdot \bm{r}} + u_{\text{rf}}(\bm{r}), \quad \bm{r} \notin \mathcal R.
\end{equation}
Where $G \in \mathbb{C}$ is the incident plane wave amplitude and $\bm{k} = (k_x, k_y, k_z)$ its wave vector, with $|\bm{k}| = k = \omega / c$ and $k_{z} > 0$. The reflected field $u_{\text{rf}}$ is complicated and has no symmetry. We  describe it in more detail in the next section.

\section{One configuration of particles}
\label{sec:one_configuration}

Solving any wave scattering from one configuration of spheres in the matrix is a difficult problem \cite{Martin_2020, Slab_book}, and as far as we know, there is no efficient semi-analytic solution for it. We formulate below the basic equations for one configuration, which we use to study the ensemble average system.

We define the spherical solutions of the Helmholtz equation $\mathrm u_{\bm{n}}$ and $\mathrm v_{\bm{n}}$ (with $\bm{n}=(\ell,m),\, \ell \in \mathbb Z_+, \, m=-\ell,\ldots,\ell$), defined as
\begin{equation}
\notag
    \begin{aligned}
        &\mathrm u_{\bm{n}}(k\bm{r}) = \mathrm u_{(\ell, m)}(k\bm{r}) :=\mathrm h_{\ell}(kr) \mathrm Y_{\bm{n}}(\hat{\bm{r}}),
\\
        &\mathrm v_{\bm{n}}(k\bm{r}) = \mathrm v_{(\ell,m)}(k\bm{r}) :=\mathrm j_{\ell}(kr) \mathrm Y_{\bm{n}}(\hat{\bm{r}}),
    \end{aligned}
\end{equation}
where $(r,\theta,\phi)$ are the spherical coordinates of $\bm{r} \in \mathbb{R}^3$; $\hat{\bm{r}}$ is the unit vector in the direction of $\bm{r}$; $\mathrm Y_{\bm{n}}(\hat{\bm{r}})$ are the spherical harmonic functions defined in Appendix \ref{app:v_to_plane}; $\mathrm j_{\ell}$ are spherical Bessel functions and $\mathrm h_{\ell}$ are spherical Hankel functions, both of the first kind.

\begin{figure}[h!]
\centering
\begin{tikzpicture}[scale = 1.5]

\draw[color = blue!10, fill=blue!10]
(-4.5,-1.8) -- (-1.8,-1.8) -- (-1.8,1.6) -- (-4.5,1.6) -- cycle;

\draw[color = yellow!40, fill=yellow!40]
(-1.8,-1.8) -- (1.6,-1.8) -- (1.6,1.6) -- (-1.8,1.6) -- cycle;
\coordinate (A) at (0,.5);
\coordinate (B) at (-1.2,0.25);
\coordinate (C) at (1,1);
\coordinate (D) at (-0.75,1.25);
\coordinate (E) at (-0.5,-0.7);
\coordinate (F) at (1.1,-0.4);
\filldraw (A) circle (2pt);
\filldraw (B) circle (3pt);
\filldraw (C) circle (2.7pt);
\filldraw (D) circle (2pt);
\filldraw (E) circle (2.6pt);
\filldraw (F) circle (3.2pt);
\draw (-1.5,-1.7) node[above]{$\mathcal{R}$};
\draw[-] (-1.8,1.6) -- (-1.8,-1.8);

\node at (-3.5,-0.5) {\textcolor{blue}{$G$}};
\node at (0,-1.3) {\textcolor{green!60!black}{$g_{\bm{n}}$}, \textcolor{red}{$f_{\bm{n}}^i$}};

\draw[->, magenta!100, thick] (-1.8,0.8)--(-2.4,1.165);
\draw[->, magenta!100, thick] (-1.8,-1.165)--(-2.2,-1.6);
\draw[->, magenta!100, thick] (-1.8,0.25)--(-2.2,0.25);
\draw[->, magenta!100, thick] (-1.8,-0.465)--(-2.3,-0.6);

\draw[<-, blue, thick] (-1.8,0.8)--(-4,0.265);
\draw[<-, blue, thick] (-1.8,-1.165)--(-4,-1.7);

\draw[<->, red, thick] (-0.58,-0.58)--(-1.12,0.12);
\draw[<->, red, thick] (-1.15,0.4)--(-0.79,1.15);
\draw[<->, red, thick] (-0.35,-0.68)--(0.95,-0.42);
\draw[<->, red, thick] (0.96,-0.28)--(0.08,0.42);
\draw[<->, red, thick] (-0.05,.4)--(-0.45,-0.57);
\draw[<->, red, thick] (0.86,1.02)--(-0.65,1.24);

\draw[<->, green!60!black, thick] (-1.8,0.25)--(-1.35,0.25);
\draw[<-, green!60!black, thick] (-0.66,-0.74)--(-1.8,-1.165);
\draw[-, green!60!black, thick] (-1.8,0.8)--(-1.3,0.35);
\draw[<-, green!60!black, thick] (-0.85,1.25)--(-1.8,0.8);
\draw[->, green!60!black, thick] (-1.8,-0.465)--(-1.28,0.1);
\draw[-, green!60!black, thick] (-1.8,-0.465)--(-0.66,-0.65);

\end{tikzpicture}
\caption{Shows the different waves that are scattered from, and arrive at, the particles and the boundary of $\mathcal R$. Each arrow identifies where the wave was generated and what it excites. The colour of each arrow indicates which term of equations \eqref{eq:total_field_medium} and \eqref{eq:total_field_hs} it is associated with. Magenta arrows represent the reflected field outside $\mathcal R$.}
\label{fig:multiple_scattering}
\end{figure}
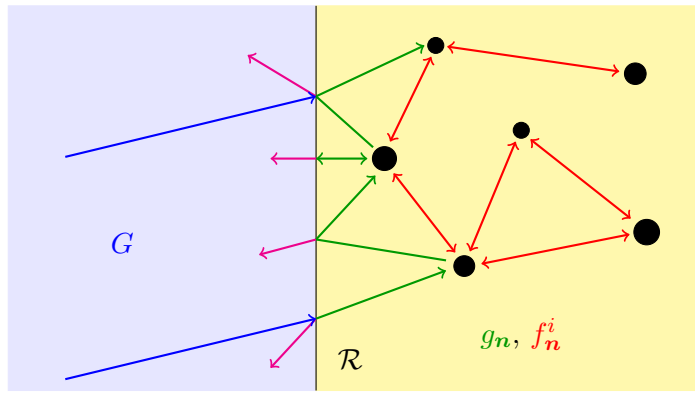

Within the matrix, and outside of the particles, the field can be written as a regular wave plus the waves scattered from each particle $u_{{\text{sc}}}^i$ in the form\footnote{The procedure to perform the summation over the double index $\bm{n} = (\ell, m)$ is given in Appendix \ref{app:add_trans}.}
\begin{align}
\label{eq:total_field_hs}
        u_{\text{tot}}(\bm{r}) &= u_{\text{reg}}(\bm{r}) + \sum_{i=1}^J u_{{\text{sc}}}^i(\bm{r}) \notag
\\ 
        &= \sum_{\bm{n}} g_{\bm{n}} \mathrm v_{\bm{n}}(k_0 \bm{r}) + \sum_{i=1}^J \sum_{\bm{n}}f_{\bm{n}}^i \mathrm u_{\bm{n}}(k_0\bm{r}-k_0\bm{r}_i), \quad \text{for} \quad \bm r \in \mathcal R \setminus \mathcal P,
\end{align}
where $\bm{r}_i$ is the position of the center of the $i$-th particle. The summations over the bold index $\bm{n}$ are performed as defined in \Cref{app:add_trans}. \Cref{fig:multiple_scattering} makes use of an arrow diagram to illustrate how waves scatter for one configuration of particles in the matrix.

Two types of waves contribute to regular field $u_{\text{reg}}(\bm{r})$: 1) the waves scattered between the boundary $\partial \mathcal R$ and the particles, and 2) the transmission of the incident field into the matrix.
As $u_{\text{reg}}(\bm{r})$ contains no sources it is smooth in $\mathcal R \setminus \mathcal P$, and can therefore be expressed in terms of regular spherical waves.

We can use the boundary conditions on each of the particles to establish a relation between $g_{\bm n}$ and the $f^i_{\bm n}$. To achieve this, we define the field that excites the $i$-th particle $u_{\text{exc}}^i(\bm{r})$. This exciting wave is the sum of the waves scattered from all $j$-th particles different from the $i$-th particle  and the background regular field:
\begin{align}
\label{def:exciting_field}
    u_{\text{exc}}^i(\bm{r}) & := u_{{\text{reg}}}(\bm{r}) + \sum_{\substack{j = 1 \\ j \not = i}}^J
    u_{{\text{sc}}}^j(\bm{r}) \notag
\\
    & = \sum_{\bm{n}} g_{\bm{n}} \mathrm v_{\bm{n}}(k_0\bm{r}) + \sum_{\bm{n}}\sum_{\substack{j \not = i}} f_{\bm{n}}^{j} \mathrm u_{\bm{n}}(k_0\bm{r} - k_0 \bm{r}_j).
\end{align}

To apply the boundary conditions for particle $i$, we write \eqref{def:exciting_field} in a basis of spherical waves centred at $\bm{r}_i$ by using \eqref{def:spherical_trans_matrices} which leads to
\begin{equation}
\label{eq:particles_boundary_conditions}
    u^i_\text{exc}(\bm r) = \sum_{\bm{nn}'}\big[ g_{\bm{n}} \mathcal{V}_{\bm{n}\bm{n}'}(k_0\bm{r}_i) + \sum_{j \not=i} f_{\bm{n}}^{j} \mathcal{U}_{\bm{nn}'}(k_0\bm{r}_i -k_0\bm{r}_j)\big]\mathrm v_{\bm{n}'}(k_0\bm{r} - k_0\bm{r}_i),
\end{equation}
for $|\bm{r} - \bm{r}_i| < |\bm{r}_j - \bm{r}_i|$. 
Solving the boundary condition for particle $i$ is now equivalent to applying the T-matrix \cite{Waterman1971,Varadan1978,MISHCHENKO1996} $T_{\bm{n}}$ to the terms multiplying $\mathrm v_{\bm{n}'}(k_0 \bm r - k_0 \bm r_i)$ in~\eqref{eq:particles_boundary_conditions}, which leads to
\begin{equation}
\label{eq:gov_eq_finite_inside}
f_{\bm{n}}^i = T_{\bm{n}} \sum_{\bm{n}'} g_{\bm{n}'} \mathcal{V}_{\bm{n}'\bm{n}}(k \bm{r}_i) + T_{\bm{n}}\sum_{j\neq i}  \sum_{\bm{n}'} f_{\bm{n}'}^{j} \, \mathcal{U}_{\bm{n}'\bm{n}}(k_0\bm{r}_i-k_0\bm r_j),
\end{equation}
where the expression of the T-matrix for a homogeneous spherical particle is given by
\begin{equation}
    \notag
    T_{\bm{n}} = T_{(\ell, m)} = - \frac{\gamma_s \mathrm j_\ell' (k_0 a) \mathrm j_\ell (k_s a) - \mathrm j_\ell (k_0 a) \mathrm j_\ell' (k_s a)}{\gamma_s \mathrm h_\ell' (k_0 a) \mathrm j_\ell (k_s a) - \mathrm h_\ell (k_0 a) \mathrm j_\ell' (k_s a)},
\end{equation}
with $\gamma_s := (\varrho_s c_s) / (\varrho_0 c_0)$ and $k_s := \omega / c_s$.

The governing equation \eqref{eq:gov_eq_finite_inside} is a straightforward generalisation of the case of particles in only one background medium. 
If the material properties of the matrix were the same as the exterior medium ($\varrho = \varrho_0$ and $c = c_0$), then $g_{\bm n}$ would represent the incident wave, as in \cite[Equation (2.7)]{gower2021effective}.

Next, we need to establish a relation between the waves inside and outside the matrix; \eqref{eq:gov_eq_finite_inside} and \eqref{eq:total_field_medium} respectively. Instead of directly using the boundary conditions~\eqref{eq:trans_bound_conditions}, it is simpler to first ensemble average the fields first. The averaging process will result in planar symmetry and simplify the form of~\eqref{eq:total_field_medium} and \eqref{eq:total_field_hs}, so that we can then apply the boundary condition at $z = 0$.

\section{Ensemble averaging}
\label{sec:ensemble}

One clear lesson from \Cref{fig:multiple_scattering} is that $g_{\bm{n}}$ and $f_{\bm{n}}^i$ each depends on the positions of all the particles. Despite the rich number of interactions for one specific configuration, we show how the average field can be simpler.

To reach the limit of an infinite number of particles, in a mathematically consistent way, we start with a cube $\mathcal R_{\alpha}^L$ with a finite number of particles $J$. In set notation we have
\begin{equation}
    \label{def:centre_region}
    \mathcal R_{\alpha}^L := \big\{ \bm{r}_i \in \mathbb{R}^3 \ | \ x_i \in (-L/2,L/2), \  y_i \in (-L/2,L/2), \  z_i \in (\alpha, L + \alpha) \big\},
\end{equation}
where $\alpha$ is chosen later to create a space between the particles and the boundary, similar to Section \ref{subsec:strat}.

Now we can consider an ensemble for particles within $\mathcal R_{\alpha}^L$. The probability density for the particles occupying the positions $\bm r_1, \bm r_2, \ldots, \bm r_J$ are given by
\begin{equation}
    \label{eq:ensemble}
    p(\bm{r}_1, \bm{r}_2, \ldots, \bm{r}_J)
\end{equation}
%
%

We define the ensemble average of any function $\mathrm f$, which implicitly depends on the position and properties of the particles, over the configuration space as
\begin{equation}
    \label{def:ensemble_average}
    \langle \mathrm f \rangle := \int_{\left( \mathcal R_{\alpha}^L \right)^{J}} \mathrm f \ p(\bm{r}_1, \bm{r}_2, \ldots, \bm{r}_J) \ \mathrm d \bm{r}_1 \mathrm d \bm{r}_2 \ldots \mathrm d \bm{r}_J,
\end{equation}
where a set to the power $J$ denotes the Cartesian product with itself $J$ times. We also define the first and second conditional ensemble averages as
\begin{equation}
    \label{def:conditional_ensemble_average}
    \begin{aligned}
        &\langle \mathrm f \rangle (\bm{r}_1) := \int_{\left( \mathcal R_{\alpha}^L \right)^{J-1}} \mathrm f \ p(\bm{r}_2, \ldots, \bm{r}_J| \bm{r}_1) \ \mathrm d \bm{r}_2 \ldots \mathrm d \bm{r}_J,
\\
        &\langle \mathrm f \rangle (\bm{r}_1, \bm{r}_2) := \int_{\left( \mathcal R_{\alpha}^L \right)^{J-2}} \mathrm f \ p(\bm{r}_3, \ldots, \bm{r}_J| \bm{r}_1,\bm{r}_2) \ \mathrm d \bm{r}_3 \ldots \mathrm d \bm{r}_J,
    \end{aligned}
\end{equation}
where we have used the marginalized probability functions for one and two particles, given by
\begin{equation}
    \label{eq:marginalized_ensembles}
    \begin{aligned}
        &p(\bm{r}_1) := \int_{\left( \mathcal R_{\alpha}^L \right)^{J-1}} p(\bm{r}_1, \ldots, \bm{r}_J) \ \mathrm d \bm{r}_2 \ldots \mathrm d \bm{r}_J,
\\
        &p(\bm{r}_1, \bm{r}_2) := \int_{\left( \mathcal R_{\alpha}^L \right)^{J-2}} p(\bm{r}_1, \ldots, \bm{r}_J) \ \mathrm d \bm{r}_3 \ldots \mathrm d \bm{r}_J.
    \end{aligned}
\end{equation}
%
%

As in \Cref{subsec:strat}, we assume particles are distributed homogeneously, which implies that
\begin{equation}
    \label{eq:uniform}
    p(\bm{r}_i) = \frac{1}{L^3} = \frac{\mathfrak n}{J}, \quad \text{with} \quad \mathfrak n := \frac{J}{L^3}.
\end{equation}
We call $\mathfrak n$ the particle number density.
%
%

For simplicity, we assume our particles are hard spheres (non-overlapping), and make use the approximation known as hole correction:
\begin{equation}
    \label{eq:hole_correction}
    p(\bm{r}_i|\bm{r}_j) =
    \left\{
    \begin{aligned}
        p(\bm{r}_i)\frac{J}{J -1} &, \quad \text{for } |\bm{r}_i - \bm{r}_j| > 2a,
\\
        0 \quad \quad \quad \quad &, \quad \text{for } |\bm{r}_i - \bm{r}_j| \leq 2a,
    \end{aligned}
    \right.
\end{equation}
where the factor $J/ (J -1 )$ comes from the fact that there are $J$  particles within the cube $\mathcal{R}_\alpha^L$. The need to add this extra factor in the case of a finite number of particles is explained in \cite[Equation (8.1.2)]{kong2004scattering}.

With our choice of pair correlation \eqref{eq:hole_correction}, and assuming the volume of $\mathcal R_{\alpha}^L$ is much larger than the volume of the particles, we conclude that the position of just one of the particles does not significantly affect the probability distribution of the other particles. In other words, we assume
\begin{align}
    \label{eq:one_particle_does_not_matter_1}
    & p(\bm{r}_2, \ldots, \bm{r}_J | \bm{r}_1) \approx p(\bm{r}_2, \ldots, \bm{r}_J),
\\
    \label{eq:one_particle_does_not_matter_2}
    & p(\bm{r}_3, \ldots, \bm{r}_J | \bm{r}_2,\bm{r}_1) \approx p(\bm{r}_3, \ldots, \bm{r}_J | \bm{r}_2).
\end{align}
Another way of interpreting these approximations is by replacing conditional probabilities with its average over $\bm{r}_1$
\begin{equation}
    \notag
    \begin{aligned}
        & p(\bm{r}_2, \ldots , \bm{r}_J| \bm{r}_1) \approx \int_{\mathcal R_{\alpha}^L} p(\bm{r}_2, \ldots, \bm{r}_J| \bm{r}_1) p(\bm{r}_1) \mathrm d \bm{r}_1 = p(\bm{r}_2, \ldots, \bm{r}_J),
\\
        & p(\bm{r}_3, \ldots, \bm{r}_J| \bm{r}_2,\bm{r}_1) \approx \int_{\mathcal R_{\alpha}^L} p(\bm{r}_3, \ldots, \bm{r}_J | \bm{r}_2,\bm{r}_1) p(\bm{r}_1|\bm{r}_2) \mathrm d \bm{r}_1 = p(\bm{r}_3, \ldots, \bm{r}_J | \bm{r}_2).
\end{aligned}
\end{equation}

\section{Average fields}
\label{sec:ensem_avrg_fields}

Our goal in this section is to describe the average scattered field close to the interface $z = 0$, so we can apply boundary conditions for the average total field. To make sure the particles do not touch the boundary at $z = 0$, as done in \Cref{subsec:strat}, we take $\alpha = a + \delta$ in \eqref{def:centre_region}, see \Cref{fig:halfspace}a. We remind the reader that $a$ is the radius of the particles and $\delta/a \ll 1$ is a small parameter.

We start by computing the average of the total field. We multiply equation \eqref{eq:total_field_medium} by \eqref{eq:ensemble}, use definition \eqref{def:ensemble_average} for some fixed value of $L$, and then integrate over all possible particle positions to obtain
\begin{equation}
    \label{eq:average_field_out}
    \ensem{u_{\text{tot}}(\bm{r})} = 
        G \mathrm e^{\mathrm i \bm k \cdot \bm r}  + \langle u_{\text{rf}} (\bm{r}) \rangle, \quad \text{for} \quad \bm{r} \notin \mathcal R,
\end{equation}
%
Here we have $\ensem{G} = G$ because the incident wave is the same for every configuration of particles. Computing the average of equation \eqref{eq:total_field_hs} results in
\begin{equation}
    \notag
    \langle u_{\text{tot}} (\bm{r}) \rangle = \langle u_{\text{reg}} (\bm{r}) \rangle + \sum_{j = 1}^J \langle u_{\text{sc}}^j (\bm{r}) \rangle,
\end{equation}
where
\begin{align}
    \label{eq:average_reg_field_inside}
    &\langle u_{\text{reg}} (\bm{r}) \rangle = \sum_{\bm{n}} \langle g_{\bm{n}} \rangle \mathrm v_{\bm{n}} (k_0 \bm{r}), \quad \text{for} \quad \bm{r} \in \mathcal R,
\\
    \label{eq:average_scat_field_inside}
    &\sum_{j = 1}^J \langle u_{\text{sc}}^j (\bm{r}) \rangle = \sum_{i=1}^J \sum_{\bm{n}} \int_{\mathcal R_{a + \delta}^L} \langle f_{\bm{n}}^i \rangle (\bm{r}_i) \mathrm u_{\bm{n}} (k_0 \bm{r} - \bm{r}_i) p(\bm{r}_i) \mathrm d \bm{r}_i, 
\end{align}
and equation \eqref{eq:average_scat_field_inside} above is valid for $0 \leq z \leq \delta$. This is because if $z > \delta$ then $\bm r$ could be inside a particle, in which case \eqref{eq:total_field_hs} would not be valid. In \eqref{eq:average_scat_field_inside}, $\langle f_{\bm{n}}^i \rangle (\bm{r}_i)$ is the average of $f_{\bm{n}}^i$ conditional to $\bm r_i$ as defined in \eqref{def:conditional_ensemble_average}${}_1$.

In our problem particles are indistinguishable, which enables us to use the simpler notation:
\begin{equation}
    \label{eq:indistinguishble}
    \langle f_{\bm{n}}\rangle (\bm{r}_i)  := \langle f_{\bm{n}}^i  \rangle (\bm{r}_i),
\end{equation}
for $i = 1,2,\ldots, J$. This notation makes it clear how to simplify the sums over particle indices. For example, the second term in \eqref{eq:average_scat_field_inside} combined with \eqref{eq:uniform} simplifies into
\begin{equation}
    \label{eq:average_particle_scattering}
    \sum_{i=1}^J\ensem{u_\text{sc}^i(\bm r)} = \mathfrak{n} \sum_{\bm{n}} \int_{\mathcal R_{a + \delta}^L} \langle f_{\bm{n}} \rangle (\bm{r}_1) \mathrm u_{\bm{n}} (k_0 \bm{r} - k_0 \bm{r}_1) \mathrm d \bm{r}_1.
\end{equation}

\subsection{The infinite volume limit}
\label{subsec:inf_J}

We now take the limit of $L \to \infty$ so that $\mathcal R_{a + \delta}^L$ becomes a halfspace, and compute the ensemble average for each term of the total average field, inside and outside the matrix.

\textbf{Average regular field.}
Because we assume particles are uniformly distributed,  see equation \eqref{eq:uniform}, and due to planar symmetry of the problem, the regular field evaluated in $z > 0$, shown in \eqref{eq:average_reg_field_inside}, can be represented as a plane wave after averaging:
\begin{equation}
    \notag
    \lim_{L \to \infty} \ensem{u_{\text{reg}} (\bm{r})} = \langle A \rangle \mathrm e^{\mathrm i \bm{k}_0 \cdot \bm{r}} + \langle A_- \rangle \mathrm e^{\mathrm i \bm{k}_0^- \cdot \bm{r}}, \quad \text{for } \quad z \geq 0,
\end{equation}
where $\bm{k}_0 = (k_{0x}, k_{0y}, k_{0z}) := \left( k_x, k_y, \sqrt{k_0^2 - k_x^2 - k_y^2} \right)$ and $\bm{k}_0^- := (k_{0x}, k_{0y}, -k_{0z})$ are the wavevectors\footnote{\label{note:Snell} Here we have used Snell's law to determine the components of the wavevectors for simplicity of the equations. However, this can be deduced directly from the transmission boundary conditions \eqref{eq:trans_bound_cond} in \Cref{subsec:boundary_conditions_hs} ahead.} of the plane waves. Without loss of generality, we choose $\mathrm{Im}[k_{0z}] \geq 0$ which implies that $\langle A_- \rangle = 0$ to avoid an unphysical wave. This choice also allows us to represent \eqref{eq:average_reg_field_inside} in terms of a single plane wave, given by
\begin{equation}
    \label{eq:reg_field_big}
    \lim_{L \to \infty} \ensem{u_{\text{reg}} (\bm{r})} = \langle A \rangle \mathrm e^{\mathrm i \bm{k}_0 \cdot \bm{r}}.
\end{equation}
The explicit expression for $\langle A \rangle$ in terms of $\langle g_{\bm{n}} \rangle$ is given in Appendix \ref{app:v_to_plane}.

\textbf{Average backscattered field.}
Taking the limit of $L \to \infty$ for the average backscattered field \eqref{eq:average_particle_scattering} leads to
\begin{equation}
\label{eq:average_particle_scattering_inf_J}
    \lim_{\substack{J \to \infty \\ L \to \infty}} \sum_{i=1}^J \ensem{u_\text{sc}^i(\bm r)} = \mathfrak{n} \sum_{\bm{n}} \int_{\mathcal R_{a + \delta}} \langle f_{\bm{n}} \rangle (\bm{r}_1) \mathrm u_{\bm{n}} (k_0 \bm{r} - k_0 \bm{r}_1) \mathrm d \bm{r}_1,
\end{equation}
where $\mathcal R_{a + \delta} := \lim_{L \to \infty} \mathcal R_{a + \delta}^L$ is the halfspace $z \geq a + \delta$. When taking this limit we need to fix the particle number density $\mathfrak n$, given in \eqref{eq:uniform}${}_2$, which implies that $J \to \infty$ when $L \to \infty$.
%

Less obviously, \eqref{eq:average_particle_scattering_inf_J} also has a plane wave representation due to symmetry. In \Cref{app:translation,app:planar_to_u} we show how to rewrite \eqref{eq:average_particle_scattering_inf_J} as
\begin{equation}
    \label{eq:average_particle_scattering_B}
    \lim_{L \to \infty} \sum_{i=1}^{\infty} \ensem{u_\text{sc}^i(\bm r)} = \langle B \rangle \mathrm e^{\mathrm i \bm{k}_0^- \cdot \bm{r}}, \quad \text{for} \quad 0 \leq z \leq \delta,
\end{equation}
where the average backscattering amplitude $\langle B \rangle$ given by
\begin{equation}
    \label{eq:average_scattered_coeff_B}
    \langle B \rangle := \frac{2\pi \mathfrak n}{k_0 k_{0z}} \sum_{\bm{n}} \mathrm i^\ell \mathrm Y_{\bm n} (\hat{\bm{k}}_0) \int_{a + \delta}^{\infty} \langle f_{\bm{n}} \rangle (0,0,z_1) \mathrm e^{\mathrm i k_{0z} z_1} \mathrm d z_1.
\end{equation}

\textbf{Average external field.}
By the same symmetry arguments, we know that the waves outside $\mathcal R$ can also be represented in terms of plane waves\textsuperscript{\ref{note:Snell}}:
\begin{equation}
    \label{eq:average_field_outside} 
    \langle u_{\text{tot}} (\bm{r}) \rangle = G \mathrm e^{\mathrm i \bm k \cdot \bm r} + \langle R \rangle \mathrm e^{\mathrm i (k_x x + k_y y - k_z z)} , \quad \text{for }\quad z \leq 0
\end{equation}
where $\langle R \rangle$ is the average reflection amplitude. In the next section, we will derive a system of equations relating the amplitudes of the different fields. This will enable us to deduce $\langle R \rangle$, which is the quantity of main interest in this paper.

\subsection{Boundary conditions at the interface}
\label{subsec:boundary_conditions_hs}

To write down relations between the average amplitudes $\ensem{A}$, $\ensem{B}$ and $\ensem{R}$, we need to impose transmission boundary conditions at $z = 0$, which reads
\begin{equation}
\label{eq:trans_bound_cond}
    \left\{
    \begin{aligned}
        & \langle u_{\text{tot}} (\bm{r}) \rangle \text{ is continuous at } z = 0,
\\
        &\frac{1}{\rho(\bm{r})} \frac{\partial \langle u_{\text{tot}} (\bm{r}) \rangle}{\partial z} \text{ is continuous at } z = 0,
    \end{aligned}
    \right.
\end{equation}
where $\rho(\bm{r})$ is the density, which is a function of $\bm r \in \mathbb R^3$.

We substitute \eqref{eq:reg_field_big}, \eqref{eq:average_particle_scattering_B} and \eqref{eq:average_field_outside} into \eqref{eq:trans_bound_cond}$_{1}$  at the boundary $\overline{\bm{r}} = (x, y, 0)$:
\begin{equation}
\notag
        G \mathrm e^{\mathrm i \bm{k} \cdot \overline{\bm{r}}} + \langle R \rangle  \mathrm e^{\mathrm i \bm{k} \cdot \overline{\bm{r}}} = \langle A \rangle \mathrm e^{\mathrm i \bm{k}_0 \cdot \overline{\bm{r}}} + \langle B\rangle \,\mathrm e^{\mathrm i \bm{k}_0 \cdot \overline{\bm{r}}}
\end{equation}
%
Because $\bm{k} \cdot \overline{\bm{r}} = \bm{k}_0 \cdot \overline{\bm{r}} = k_x x + k_y y$, the above simplifies to
\begin{equation}
    \label{eq:amplitudes_system_eq1}
    G + \langle R \rangle = \langle A \rangle + \langle B \rangle.
\end{equation}
%
Similar computations can be done as above, starting from \eqref{eq:trans_bound_cond}$_{2}$ instead of \eqref{eq:trans_bound_cond}$_{1}$. These result in another relation,
\begin{equation}
    \label{eq:amplitudes_system_eq2}
     \frac{k_z}{\varrho} \left( G - \langle R \rangle \right) = \frac{k_{0z}}{\varrho_0} \left( \langle A \rangle - \langle B \rangle \right).
\end{equation}
%
The system of equations \eqref{eq:amplitudes_system_eq1}-\eqref{eq:amplitudes_system_eq2} can be rearranged into
\begin{equation}
    \label{eq:sol_bound_cond_hs}
        \langle R \rangle = \zeta_R G + \zeta_T \langle B \rangle, \quad
         \langle A \rangle = \gamma_0 \zeta_T G - \zeta_R \langle B \rangle,
\end{equation}
where the newly introduced parameters $\zeta_R,\,\zeta_T,\,\gamma_0$ are given by
\begin{equation}
    \notag
    \zeta_R := \frac{\varrho_0 \, k_z - \varrho \, k_{0z}}{\varrho_0 \, k_z + \varrho \, k_{0z}},
\quad
    \zeta_T := \frac{2 \varrho \, k_{0z}}{\varrho_0 \, k_z + \varrho \, k_{0z}},
\quad
    \gamma_0 := \frac{\varrho_0 \, k_z}{\varrho \, k_{0z}}.
\end{equation}
At this point, we have three unknowns $\ensem{A}$, $\ensem{B}$, and $\ensem{R}$ and two equations \eqref{eq:sol_bound_cond_hs}. The final equation comes from how the particles reflect waves combined with the expression \eqref{eq:average_scattered_coeff_B} for $\ensem{B}$. 

\subsection{Single medium limit}
\label{subsec:one_medium}


As a sanity check, we can see how the equations above recover the single background medium limit. Taking the acoustic properties of the exterior medium and the matrix as the same ($c_0 = c$ and $\varrho_0 = \varrho$), we have that $\bm{k}_0 = \bm{k}$, which means $\zeta_R = 0$ and $\zeta_T = \gamma_0 = 1$. Substituting these values in \eqref{eq:sol_bound_cond_hs} provides us with 
\begin{equation}
    \notag
       \langle A \rangle = G,  \quad  \langle R \rangle = \langle B \rangle = \frac{2\pi \mathfrak n}{k k_{z}} \sum_{\bm{n}} \mathrm i^\ell \mathrm Y_{\bm n} (\hat{\bm{k}}) \int_{a}^{\infty} \langle f_{\bm{n}} \rangle (z_1) \mathrm e^{\mathrm i k z_1} \mathrm d z_1,
\end{equation}
where we used \eqref{eq:average_scattered_coeff_B} and took the limit $\delta \to 0$ to recover the formulas for particles distributed in a halspace region $\mathcal R$, see \Cref{fig:halfspace}b. The above is the same formula for reflection of a halfspace as \cite[Eq. (7.6)]{gower2021effective}. This means the approach is consistent with the average response of random particulate materials in the case of particles embedded in only one homogeneous medium.

\section{Average backscattering operator}
\label{sec:backscattering-operator}


To obtain the final equation needed to determine the average amplitudes in \eqref{eq:sol_bound_cond_hs}, we follow the notation introduced in \Cref{subsec:strat}. That is, we need to find an equation relating $\ensem{B}$ and $\ensem{A}$ through some backscattering operator $\ensem{\mathbb T_\sigma}$. This final equation comes from the microstructure, which in our case is the  ensemble averaged version of the boundary conditions of the particles~\eqref{eq:gov_eq_finite_inside}.

Following the strategy of \cite{gower2021effective}, we take a conditional ensemble average of~\eqref{eq:gov_eq_finite_inside}, and use assumptions~\eqref{eq:uniform}, \eqref{eq:hole_correction}, \eqref{eq:indistinguishble} to obtain
%
%
%
\begin{equation}
    \begin{aligned}
    \label{eq:gov_eq_inside_average}
        \ensem{f_{\bm{n}}}(\bm{r}_1) = & \, T_{\bm{n}} \sum_{\bm{n}'} \ensem{g_{\bm{n}'}} (\bm{r}_1) \mathcal{V}_{\bm{n}'\bm{n}}(k_0 \bm{r}_1)
\\ 
        &+ \mathfrak{n} \, T_{\bm{n}} \sum_{\bm{n}'} \int_{\mathcal R_a \backslash \mathcal{B}(\bm{r}_1, 2a)} \mathcal{U}_{\bm{n}'\bm{n}}(k_0\bm{r}_1 - k_0\bm{r}_2) \langle f_{\bm{n}'} \rangle (\bm{r}_2) \mathrm d\bm{r}_2.
    \end{aligned}
\end{equation}
where we used the standard Quasi-Crystalline Approximation (QCA) to substitute $\ensem{f_{\bm{n}} }(\bm{r}_2,\bm{r}_1) = \ensem{f_{\bm{n}} } (\bm{r}_2)$ \cite{gower2021effective}.

By assuming that \eqref{eq:gov_eq_inside_average} has a unique solution for $\langle f_{\bm{n}} \rangle (\bm{r}_1)$ given $\langle g_{\bm{n}'} \rangle (\bm{r}_1)$, then, formally, we can use \eqref{eq:gov_eq_inside_average} to represent $\langle f_{\bm{n}} \rangle (\bm{r}_1)$ as
\begin{equation}
    \label{eq:formal_operator}
    \langle f_{\bm{n}} \rangle (\bm{r}_1) = \sum_{\bm{n}'} \mathcal L_{\bm{n} \bm{n}'}^f [\langle g_{\bm{n'}} \rangle (\bm{r}_1')],
\end{equation}
%
for some linear operator $\mathcal L_{\bm{n} \bm{n}'}^f$ acting on $\langle g_{\bm{n'}} \rangle (\bm{r}_1')$. This operator notation  $\mathcal L_{\bm{n} \bm{n}'}^f$ will help to compute $\ensem{\mathbb T_\sigma}$.
From \eqref{eq:average_scattered_coeff_B} we see that $\langle B \rangle$ is a linear map acting on $\ensem{f_{\bm{n}}}$. For convenience, let us also represent \eqref{eq:average_scattered_coeff_B} in the form  $\langle B \rangle = \sum_{\bm{n}} \mathcal L_{\bm{n}}^B [\langle f_{\bm{n}} \rangle (\bm{r}_1)]$. Substituting \eqref{eq:formal_operator} into this representation leads to
 %
 \begin{equation}
    \label{eq:operator_before_X-QCA}
     \langle B \rangle 
     = \sum_{\bm{n}\bm{n}'}  \mathcal L_{\bm{n}}^B \left[ \mathcal L^f_{\bm{n} \bm{n}'} [\langle g_{\bm{n}'} \rangle (\bm{r}_1')] \right].
 \end{equation}

It is not obvious, but the above is the same as \eqref{eq:system_avrg_2}. And, just like in \Cref{subsec:strat}, when combining the above with the boundary conditions \eqref{eq:sol_bound_cond_hs}, we still have to many unknowns and require a closure assumption. The closure assumption which is consistent with the standard QCA is given by  
\begin{equation} \label{eq:xqca-approximation}
\tcboxmath{
    \begin{aligned}
    & \ensem{f_{\bm{n}} }(\bm{r}_2,\bm{r}_1) = \ensem{f_{\bm{n}} } (\bm{r}_2),
    \\
    & \ensem{g_{\bm{n}}} (\bm{r}_1) = \ensem{g_{\bm{n}}}.
    \end{aligned}
    \quad \text{(X-QCA)},
    }
\end{equation} 
where X-QCA stands for the eXtended Quasi-Crystalline Approximation. Below we show how this matches the closure assumption \eqref{eq:approx_interface} in the introduction. In \Cref{sec:qca} we deduce X-QCA from first principles, and show why it is the consistent way to extend QCA. 

%

Substituting \eqref{eq:xqca-approximation}${}_2$ into \eqref{eq:operator_before_X-QCA}, and using \eqref{def:C_coefs}, leads to
\begin{equation}
    \notag
    \langle B \rangle = \langle \mathbb T_{\sigma} \rangle \langle A \rangle, \quad \text{with} \quad 
    \langle \mathbb T_{\sigma} \rangle \equiv \sum_{\bm{n}\bm{n}'}  \mathcal L_{\bm{n}}^B \left[ \mathcal L^f_{\bm{n} \bm{n}'} [\mathrm C_{\bm{n}'}] \right], 
\end{equation}
where the coefficients $\mathrm C_{\bm{n}'}$ are known quantities defined in \Cref{app:v_to_plane}.
We clearly see now how the above is equivalent to approximation \eqref{eq:approx_interface} in \Cref{subsec:strat}. 

With the above, together with \eqref{eq:sol_bound_cond_hs}, we can now calculate the solution, with an efficient numerical scheme shown in \Cref{sec:effective_waves}, where we use the Effective waves method. To summarise, we can now obtain a governing equation for the unknowns by substituting \eqref{eq:operator_before_X-QCA} into \eqref{eq:gov_eq_inside_average}, and using \eqref{def:C_coefs}, to obtain:
%
\begin{equation}
    \label{eq:gov_eq_inside_average_qca}
    \tcboxmath{
    \begin{aligned}
        \ensem{f_{\bm{n}}}(\bm{r}_1) = & \, \ensem{A} T_{\bm{n}} \sum_{\bm{n}'} \mathrm C_{\bm{n}'}\mathcal{V}_{\bm{n}'\bm{n}}(k_0 \bm{r}_1)
\\ 
        & + \mathfrak{n} \, T_{\bm{n}} \sum_{\bm{n}'} \int_{\mathcal R_a \backslash \mathcal{B}(\bm{r}_1, 2a)} \mathcal{U}_{\bm{n}'\bm{n}}(k_0\bm{r}_1 - k_0\bm{r}_2) \langle f_{\bm{n}'} \rangle (\bm{r}_2) \mathrm d\bm{r}_2.
    \end{aligned}    
    }
\end{equation}
We could now, in theory, numerically solve for $\ensem{f_{\bm{n}}}(\bm{r}_1)$, $\ensem{A}$, $\ensem{B}$, and $\ensem{R}$, by combining the above with the boundary conditions \eqref{eq:sol_bound_cond_hs} and the definition of $\ensem{B}$ in terms of the $\ensem{f_{\bm{n}}}(\bm{r}_1)$ given by \eqref{eq:average_scattered_coeff_B}. However, we present a far more efficient method to solve this in \Cref{sec:effective_waves}.

\subsection{The average reflection coefficient}
\label{sec:final_average_gov_eqs}

With all the computations so far we have successfully deduced the contribution to sound wave scattering due to particles \eqref{eq:gov_eq_inside_average_qca} and the halfspace interface \eqref{eq:sol_bound_cond_hs} as two separate equations, as briefed in \Cref{subsec:strat}. Then, we substitute \eqref{eq:sol_bound_cond_hs}$_{(2)}$ into \eqref{eq:gov_eq_inside_average_qca} to compute a single integral equation in $\langle f_{\bm{n}} \rangle (\bm{r}_1)$, which reads
\begin{equation}
    \label{eq:final_gov_eq}
    \begin{aligned}
        \ensem{f_{\bm{n}}} (\bm{r}_1) =& \ \gamma_0 \zeta_{T} \, T_{\bm{n}} W_{\bm{n}} (\bm{r}_1) G + T_{\bm{n}} D_{\bm{n}}(\bm{r}_1) \left[ \ensem{f_{\bm{n}'}} \right] - \zeta_R \, T_{\bm{n}} W_{\bm{n}} (\bm{r}_1) \langle B \rangle [\langle f_{\bm{n}'} \rangle],
    \end{aligned}
\end{equation}
where we have defined the particle-rescattering operator
\begin{equation}
    \notag
    D_{\bm{n}}(\bm{r}) [\langle f_{\bm{n}'} \rangle] := \mathfrak n \sum_{\bm{n}'} \int_{\mathcal{R}_a \backslash \mathcal{B}(\bm{r}, 2a)} \mathcal{U}_{\bm{n}'\bm{n}}(k_0\bm{r} - k_0\bm{r}') \langle f_{\bm{n}'} \rangle(\bm{r}') \mathrm d\bm{r}'.
\end{equation}
and the interface coupling function
\begin{equation}
    \label{def:interface_coupling}
    W_{\bm{n}}(\bm{r}_1) := \sum_{\bm{n}'} \mathrm C_{\bm{n}'} \mathcal{V}_{\bm{n}' \bm{n}}(k_0 \bm{r}_1) = \mathrm C_{\bm{n}} \mathrm e^{\mathrm i \bm{k}_0 \cdot \bm{r}_1},
\end{equation}
where we used \cite[eqs. (A.2) and (B.12)]{gower2021effective} to write the second equality in \eqref{def:interface_coupling}.

If the geometry of the matrix containing the particles is changed, the same equation \eqref{eq:final_gov_eq} can still be used, however, the terms involving the interface coupling factor \eqref{def:interface_coupling} will change. Also, the explicit equation for the boundary conditions \eqref{eq:sol_bound_cond_hs} will not be the same.


As we shall see further on, equation \eqref{eq:final_gov_eq} is enough to determine $\langle f_{\bm{n}} \rangle (\bm{r}_1)$, making it possible to compute the expression of the total wave outside the halfspace $\mathcal R$. We recall the expression for the reflection coefficient \eqref{eq:sol_bound_cond_hs}$_{(1)}$ below: 
\begin{equation}
\notag
     \tcboxmath{
     \ensem{R} =  \zeta_R \, G + \zeta_T \, \langle B \rangle.
     }
\end{equation}
%
The first term is the reflection coefficient from a homogeneous matrix without any particles inside. Only the second term $\zeta_T \, \langle B \rangle$ 
%
%
carries the effects of scattering from the particles. In others words, if we knew the background material, and wanted to use a reflection experiment to characterise the particles, then we should use the term $\ensem{R} -  \zeta_R \, G $ to do so.



There have been many uses of reflection coefficients from particulates in the literature, see for example \cite{Garcia-Valenzuela:05,Chung_2011,Simon_tony2024}. In \cite{Simon_tony2024} they used approximate formulas for the reflection coefficient \eqref{eq:sol_bound_cond_hs}$_{(1)}$ and obtained a good qualitative agreement with experimental data \cite{Simon_tony2024}. Having the exact formula \eqref{eq:sol_bound_cond_hs}$_{(1)}$, calculated from first principles, would likely lead to more accurate predictions, specially for a broad frequency range, which is less understood.


\section{Extended quasi-crystalline approximation}
\label{sec:qca}

In this section, we explain why the eXtended Quasi-Crystalline Approximation (X-QCA) \eqref{eq:xqca-approximation} is accurate, for low and high volume fraction of particles, and show that it is the systematic extension of the standard Quasi-Crystalline Approximation (QCA) \cite{Lax_QCA_1952} to scenarios with different background mediums.

To justify X-QCA~\eqref{eq:xqca-approximation} it is best to start with just one configuration of particles, as this will help us understand the role of the different waves.

We recall from \Cref{fig:multiple_scattering} that the field~\eqref{def:exciting_field} which excites particle $i$ depends on the configuration of all particles, including its own position  $\bm{r}_i$. Take, for example, any wave which is initially scattered from particle $i$ and then returns to particle $i$ due to multiple scattering. 
This type of dependence is called self-interaction, and it is known in the statistical mechanics literature that accounting for self-interactions can lead to unsolvable equations, or even divergences, when ensemble averaging \cite{Hynne_1987,Mishchenko_2018}. This is the case for any particulate material, whether there are different background mediums or not. 

For clarity, let us introduce the notation:
\[
u_\text{exc}^1(\bm r; \bm r_1, \bm r_2, \ldots, \bm r_J) := u_\text{exc}^1(\bm r),
\]
to denote the exciting field $u_\text{exc}^1$, so that we can explicitly discuss how $u_\text{exc}^1$ depends on the positions $\bm r_j$ of each particle.

A possible strategy to simplify the self-interactions is to approximate the field exciting particle $1$, for example, by its own conditional average:
\begin{equation}
\label{eq:exciting_field_average}
\tcboxmath{
    u_\text{exc}^1(\bm r; \bm r_1, \bm r_2, \ldots, \bm{r}_J) \approx  \ensem{u_\text{exc}^1}_{1}(\bm r; \bm r_2, \ldots, \bm{r}_J) 
}    
\end{equation}
where we used the conditional average:
\begin{equation} \label{def:1-conditional_average}
 \ensem{u_\text{exc}^1}_{1}(\bm r; \bm r_2, \ldots, \bm{r}_J) := \int_{\mathcal{R}_{a + \delta}^L} u_\text{exc}^1(\bm{r};\bm r_1', \bm r_2, \ldots, \bm{r}_J) p(\bm{r}_1'|\bm{r}_2,\ldots, \bm{r}_J) \mathrm d \bm{r}_1'.
\end{equation}
We will show how~\eqref{eq:exciting_field_average} leads the standard QCA \cite{Lax_QCA_1952} when there is just one background medium, and it also leads to X-QCA when there is different background mediums and interfaces. But first let us consider whether it is a sensible approximation.

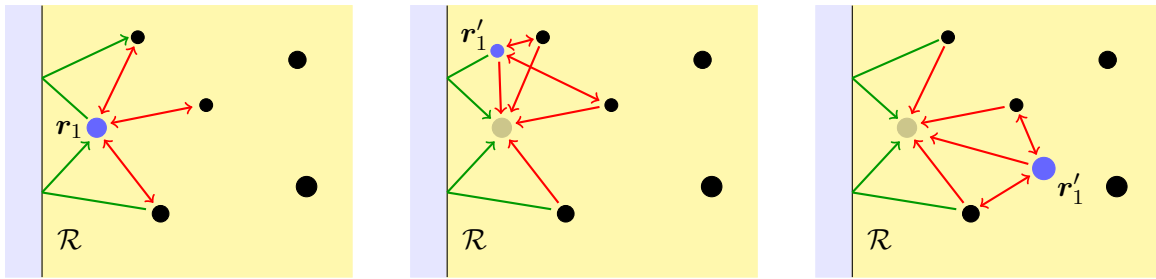
\begin{figure}[h!]
\centering
\begin{tikzpicture}[scale=1.2]
\draw[color = blue!10, fill=blue!10]
(-2.2,-1.4) -- (-1.8,-1.4) -- (-1.8,1.6) -- (-2.2,1.6) -- cycle;

\draw[color = yellow!40, fill=yellow!40]
(-1.8,-1.4) -- (1.6,-1.4) -- (1.6,1.6) -- (-1.8,1.6) -- cycle;
\coordinate (A) at (0,.5);
\coordinate (B) at (-1.2,0.25);
\coordinate (C) at (1,1);
\coordinate (D) at (-0.75,1.25);
\coordinate (E) at (-0.5,-0.7);
\coordinate (F) at (1.1,-0.4);
\filldraw (A) circle (2pt);
\filldraw (B)[blue!60] circle (3pt);
\filldraw (C) circle (2.7pt);
\filldraw (D) circle (2pt);
\filldraw (E) circle (2.6pt);
\filldraw (F) circle (3.2pt);
\draw (-1.5,-1.2) node[above]{$\mathcal{R}$};
\draw[-] (-1.8,1.6) -- (-1.8,-1.4);
\draw (-1.5,0.02) node[above]{$\bm{r}_1$};

\draw[<->, red, thick] (-0.58,-0.58)--(-1.12,0.12);
\draw[<->, red, thick] (-1.15,0.4)--(-0.79,1.15);
\draw[<->, red, thick] (-0.15,0.48)--(-1.05,0.3);

\draw[-, green!60!black, thick] (-1.8,0.8)--(-1.3,0.35);
\draw[<-, green!60!black, thick] (-0.85,1.25)--(-1.8,0.8);
\draw[->, green!60!black, thick] (-1.8,-0.465)--(-1.28,0.1);
\draw[-, green!60!black, thick] (-1.8,-0.465)--(-0.66,-0.65);
\end{tikzpicture}
\hspace{0.5cm}
\begin{tikzpicture}[scale=1.2]
\draw[color = blue!10, fill=blue!10]
(-2.2,-1.4) -- (-1.8,-1.4) -- (-1.8,1.6) -- (-2.2,1.6) -- cycle;

\draw[color = yellow!40, fill=yellow!40]
(-1.8,-1.4) -- (1.6,-1.4) -- (1.6,1.6) -- (-1.8,1.6) -- cycle;
\coordinate (A) at (0,.5);
\coordinate (B) at (-1.2,0.25);
\coordinate (C) at (1,1);
\coordinate (D) at (-0.75,1.25);
\coordinate (E) at (-0.5,-0.7);
\coordinate (F) at (1.1,-0.4);
\coordinate (G) at (-1.25,1.1);
\filldraw (A) circle (2pt);
\filldraw[opacity = 0.2] (B) circle (3pt);
\filldraw (C) circle (2.7pt);
\filldraw (D) circle (2pt);
\filldraw (E) circle (2.6pt);
\filldraw (F) circle (3.2pt);
\filldraw (G)[blue!60] circle (2pt);
\draw (-1.5,-1.2) node[above]{$\mathcal{R}$};
\draw[-] (-1.8,1.6) -- (-1.8,-1.4);
\draw (-1.5,1) node[above]{$\bm{r}_1'$};

\draw[->, red, thick] (-0.58,-0.58)--(-1.12,0.12);
\draw[<-, red, thick] (-1.1,0.42)--(-0.79,1.15);
\draw[->, red, thick] (-0.13,0.48)--(-1.05,0.3);
\draw[->, red, thick] (-1.23,0.98)--(-1.21,0.4);
\draw[<->, red, thick] (-1.15,1.15)--(-0.85,1.23);
\draw[<->, red, thick] (-1.15,1.05)--(-0.13,0.55);

\draw[->, green!60!black, thick] (-1.8,0.8)--(-1.32,0.35);
\draw[-, green!60!black, thick] (-1.35,1.05)--(-1.8,0.8);
\draw[->, green!60!black, thick] (-1.8,-0.465)--(-1.28,0.1);
\draw[-, green!60!black, thick] (-1.8,-0.465)--(-0.66,-0.65);
\end{tikzpicture}
\hspace{0.5cm}
\begin{tikzpicture}[scale=1.2]
\draw[color = blue!10, fill=blue!10]
(-2.2,-1.4) -- (-1.8,-1.4) -- (-1.8,1.6) -- (-2.2,1.6) -- cycle;

\draw[color = yellow!40, fill=yellow!40]
(-1.8,-1.4) -- (1.6,-1.4) -- (1.6,1.6) -- (-1.8,1.6) -- cycle;
\coordinate (A) at (0,.5);
\coordinate (B) at (-1.2,0.25);
\coordinate (C) at (1,1);
\coordinate (D) at (-0.75,1.25);
\coordinate (E) at (-0.5,-0.7);
\coordinate (F) at (1.1,-0.4);
\coordinate (G) at (0.3,-0.2);
\filldraw (A) circle (2pt);
\filldraw[opacity = 0.2] (B) circle (3pt);
\filldraw (C) circle (2.7pt);
\filldraw (D) circle (2pt);
\filldraw (E) circle (2.6pt);
\filldraw (F) circle (3.2pt);
\filldraw (G)[blue!60] circle (3.5pt);
\draw (-1.5,-1.2) node[above]{$\mathcal{R}$};
\draw[-] (-1.8,1.6) -- (-1.8,-1.4);
\draw (0.6,-0.7) node[above]{$\bm{r}_1'$};

\draw[->, red, thick] (-0.58,-0.58)--(-1.12,0.12);
\draw[<-, red, thick] (-1.15,0.4)--(-0.79,1.15);
\draw[->, red, thick] (-0.13,0.48)--(-1.05,0.3);
\draw[<-, red, thick] (-0.95,0.15)--(0.13,-0.15);
\draw[<->, red, thick] (0.03,0.4)--(0.22,-0.05);
\draw[<->, red, thick] (0.15,-0.28)--(-0.4,-0.62);

\draw[->, green!60!black, thick] (-1.8,0.8)--(-1.3,0.35);
\draw[-, green!60!black, thick] (-0.85,1.22)--(-1.8,0.8);
\draw[->, green!60!black, thick] (-1.8,-0.465)--(-1.28,0.1);
\draw[-, green!60!black, thick] (-1.8,-0.465)--(-0.66,-0.65);

\end{tikzpicture}
\caption{The image on the left shows how waves scattered from the particle at $\bm{r}_1$ (blue disk) contribute to the field $u^1_\text{exc}(\bm r; \bm r_1; \ldots)$ evaluated for $\bm r$ close to $\bm{r}_1$. The image in the centre, and on the right, show how the exciting field  $u^1_\text{exc}(\bm r; \bm r_1'; \ldots)$, evaluated near $\bm r$, changes when moving  particle-1 to the position $\bm{r}_1'$. Note that $\bm{r}_1' = \bm{r}_1$ is always a feasible position, as there is no other particle at $\bm r_1$, and therefore the case $\bm{r}_1' = \bm{r}_1$ always has a significant contribution in the integral~\eqref{def:1-conditional_average} when calculating $\ensem{u^1_\text{exc}}_1$. }
\label{fig:quasicrystaline}
\end{figure}

Is the approximation~\eqref{eq:exciting_field_average} accurate? \Cref{fig:quasicrystaline} illustrates some of the possible positions $\bm{r}_1'$ of the first particle, and how they contribute to the field exciting the particle positioned at $\bm{r}_1$. We will explain why~\eqref{eq:exciting_field_average} is highly accurate for large and small particle volume fractions. We note that similar arguments have been used in \cite{waterman1961multiple, fikioris1964multiple,Varadans_84}.

For a small volume fraction of particles, the wave scattered from particle $i$ will be weakly rescattered by the same particle $i$, as the average distances to the next particle, or interface, are large. In which case $u_{\text{exc}}^1(\bm{r}; \bm r_1, \bm r_2, \ldots)$
weakly depends on the position of particle $1$. From this we can deduce the approximation \eqref{eq:exciting_field_average} by: taking $u_{\text{exc}}^1(\bm{r}; \bm r_1, \bm r_2, \ldots)$ out of the integral in \eqref{def:1-conditional_average} together with
\[
    \int_{\mathcal{R}_{a + \delta}^L} p(\bm{r}_1'|\bm{r}_2 \ldots, \bm{r}_J) \mathrm d \bm{r}_1' \approx 1.
\]

For a large volume fraction of particles, most of all particle positions $\bm{r}_1'$ in the integral \eqref{def:1-conditional_average} would be prohibitive, as the particles are densely packed together and particles can not overlap. That is, the function $p(\bm{r}_1'|\bm{r}_2 \ldots, \bm{r}_J)$ is zero when particles overlap. This is illustrated in \Cref{fig:quasicrystaline}. The one region that most contributes to the integral in \eqref{def:1-conditional_average} is the region around the original particle position $\bm r_1$. An extreme example of this is a crystal, where the neighbouring particles exactly determine the position of $\bm{r}_1'$ and 
\[
p(\bm{r}_1'|\bm{r}_2,\ldots, \bm{r}_J) = \delta (\bm{r}_1' - \bm{r}_1),
\] 
which substituted into~\eqref{eq:exciting_field_average} leads to 
$u_{\text{exc}}^1(\bm{r}; \bm r_1, \bm r_2 \ldots) = \ensem{u_\text{exc}^1}_{1}(\bm{r};\bm r_2, \ldots)$ exactly.

The discussion above shows why~\eqref{eq:exciting_field_average} is a good approximation, which justifies why it commonly used in the literature \cite{Varadans_84,Mishchenko_2011,Linton_2005, gower2019multiple, Slab_book,Aris2024,Kevish2024}. In the next section we show how approximation \eqref{eq:exciting_field_average} leads to X-QCA \eqref{eq:xqca-approximation}, and therefore is equivalent to standard QCA when there is just one background medium.

\subsection{The average exciting field}
\label{subsec:particle_average}

To reach X-QCA \eqref{eq:xqca-approximation} (or the standard QCA) from the approximation~\eqref{eq:exciting_field_average} we: start with the definition~\eqref{def:exciting_field}, rename $\bm{r}_1$ to $\bm{r}_1'$, multiply both sides by $p(\bm{r}_1'|\bm{r}_2,\ldots, \bm{r}_J)$, integrate over $\bm{r}_1'$, and then use~\eqref{def:1-conditional_average} together with the approximation \eqref{eq:exciting_field_average} to arrive at 
\begin{equation}
\label{eq:explicit_average}
    u^1_\text{exc}(\bm{r}) = \sum_{\bm{n}} \ensem{g_{\bm{n}}}_1 \mathrm v_{\bm{n}}(k_0\bm{r}) + \sum_{\bm{n}}\sum_{j=2}^J \ensem{f_{\bm{n}}^{j}}_1 \mathrm u_{\bm{n}}(k_0\bm{r} - k_0 \bm{r}_j).
\end{equation}
Note, in practice the only difference between the above and~\eqref{def:exciting_field} is that we have replaced $g_{\bm n}$ by $\ensem{g_{\bm n}}_1$ and $f_{\bm n}^j$ by $\ensem{f_{\bm n}^j}_1$ for $j \not = 1$.

Following the same steps as done in \Cref{sec:ensem_avrg_fields,sec:backscattering-operator}, we would obtain the same expression as in \eqref{eq:gov_eq_inside_average}, except with the substitutions
\[
\ensem{g_{\bm n}}(\bm r_1) = \ensem{\ensem{g_{\bm n}}_1}(\bm r_1) \;\; \text{and} \;\; \ensem{f_{\bm n}}(\bm r_2, \bm r_1) = \ensem{\ensem{f_{\bm n}}_1}(\bm r_2, \bm r_1),
\]
we show below that these are equivalent to~\eqref{eq:xqca-approximation} for disordered particulates.
Combining the definition~\eqref{def:1-conditional_average} (with $g_n$ in place of $ u^1_\text{exc}$) with the definition~\eqref{def:conditional_ensemble_average} leads to
\begin{equation}
    \label{eq:average_reg_coef}
    \begin{aligned}
        \ensem{\ensem{g_{\bm{n}}}_1}(\bm{r}_1) 
         =& \int g_{\bm{n}} p(\bm{r}_1'|\bm{r}_2, \ldots) p(\bm{r}_2, \ldots|\bm{r}_1) \mathrm d \bm{r}_1' \mathrm d \bm{r}_2 \ldots \mathrm d \bm{r}_J
\\
        =& \int g_{\bm{n}} \frac{p(\bm{r}_1', \bm{r}_2, \ldots)}{p(\bm{r}_2, \bm{r}_3, \ldots)} p(\bm{r}_2,\bm{r}_3, \ldots |\bm{r}_1) \mathrm d \bm{r}_1' \mathrm d \bm{r}_2 \ldots \mathrm d \bm{r}_J
\\
        =& \int g_{\bm{n}} 
        p(\bm{r}_1', \bm{r}_2, \ldots) \mathrm d \bm{r}_1' \ldots \mathrm d \bm{r}_J
        = \langle g_{\bm{n}} \rangle,
    \end{aligned}
\end{equation}
where in the third line we used assumption \eqref{eq:one_particle_does_not_matter_1}, which is valid for a large number of disordered particles. Similarly, we have that 
\begin{equation}
    \label{eq:average_scat_coef}
    \begin{aligned}
        \langle \ensem{f_{\bm{n}'}^{2}}_1 \rangle (\bm{r}_2,  \bm{r}_1) =& \int f_{\bm{n}'}^2 p(\bm{r}_1'|\bm{r}_2,\ldots) p(\bm{r}_3, \ldots|\bm{r}_2, \bm{r}_1) \mathrm d \bm{r}_1' \mathrm d \bm{r}_3 \ldots \mathrm d \bm{r}_J
\\
        =& \int f_{\bm{n}'}^2 \frac{p(\bm{r}_1', \ldots)}{p(\bm{r}_2, \ldots)} p(\bm{r}_3, \ldots |\bm{r}_2, \bm{r}_1) \mathrm d \bm{r}_1' \mathrm d \bm{r}_3 \ldots \mathrm d \bm{r}_J
\\
        =& \int f_{\bm{n}'}^2 
        \frac{p(\bm{r}_1', \bm{r}_3, \ldots |\bm{r}_2)}{p(\bm{r}_3.,\ldots|\bm{r}_2)} 
        p(\bm{r}_3, \ldots|\bm{r}_2, \bm{r}_1) \mathrm d \bm{r}_1' \mathrm d \bm{r}_3 \ldots \mathrm d \bm{r}_J
\\
        =& \int f_{\bm{n}'}^2 p(\bm{r}_1', \bm{r}_3, \ldots |\bm{r}_2) \mathrm d \bm{r}_1' \mathrm d \bm{r}_3 \ldots \mathrm d \bm{r}_J
\\
        =& \langle f_{\bm{n}'}^2 \rangle (\bm{r}_2) = \langle f_{\bm{n}'} \rangle (\bm{r}_2),
    \end{aligned}
\end{equation}
where we repeatedly used the definition of conditional probability, and in the last but one line we used assumption \eqref{eq:one_particle_does_not_matter_2}. 

The results above demonstrate that approximation~\eqref{eq:xqca-approximation} is a consequence of approximation~\eqref{eq:exciting_field_average} for disordered particulates as defined in \Cref{sec:ensemble}. Because the approximation holds for any number of particles and any size of $|\mathcal R_{a+\delta}^L|$, \eqref{eq:xqca-approximation} is valid for the limit of infinite particles defined in \Cref{subsec:inf_J}.

An advantage of using approximations in terms of the exciting field \eqref{eq:exciting_field_average}, instead of quantities which are more directly related to the particles, such as \eqref{eq:xqca-approximation}, is that it is clear how to extend this approximation to more complex scenarios. In the presence of other geometries, layers, or multispecies \cite{gower_reflection_2018}, we can still define an exciting field, and then use \eqref{eq:exciting_field_average} directly. It is also possible to account for more complex interactions between particles other than the pair correlation \eqref{eq:hole_correction} assumed in this work. See \cite{twersky1964,Mishchenko_2011} for a broader discussion on how to account for different pair-correlations for standard QCA. In other words, approximation \eqref{eq:exciting_field_average} leads to a systematic way to generalise the original QCA introduced in \cite{Lax_QCA_1952}. 

Beyond generalising QCA, we feel that approximation \eqref{eq:exciting_field_average} provides more physical insight. We saw from the section above that QCA, and its extension, only approximate the self-interaction by averaging it (conditioned on the position of the other particles). This has already been discussed for the classical QCA through a scattering series expansion \cite{Mishchenko_2011}. As a consequence, QCA, and its extension, is only making an approximation about third-order and higher scattering\footnote{An example of third-order self-interaction scattering would be: the incident wave scatters on the $i$-th particle, which then gets scattered by the $j$-th particle, and finally reaches the $i$-th particle again, scattering once more. See Fig. 3 in \cite{Mishchenko_2011} for the Feynman diagrams.}. This sheds light on the agreement between second-order weak scattering approximations and QCA, as discussed in \cite{Martin2008}. 

Finally, we note that some work has been carried out in the literature \cite{Slab_book} which was able to write down a closed system of equations by using the classical QCA, but without making an approximation on the regular field $\ensem{g_{\bm n}}(\bm r_1)$. However, when using QCA, we would already be making an approximation about third-order scattering, so there is no reason to retain high order scattering for some terms (from a wall) but not from others (particles). Further, X-QCA leads to systems which are far simpler to solve, as we demonstrate in \Cref{sec:effective_waves}.



\section{Effective waves method}
\label{sec:effective_waves}

We use the Effective waves method, introduced in \cite{gower2021effective}, to solve the governing equation \eqref{eq:final_gov_eq}. As shown in \cite{gower2021effective}, it is usually accurate, even for a broad range of frequencies and particle volume fractions\footnote{The general solution presented in \cite{gower2021effective} introduces multiple wavenumbers, although they show that there is usually only one dominant wavenumber $k_*$ for most properties and frequencies. See \cite{Aris2024} for phase diagrams that show when more than one wavenumber is needed.} to assume that $\ensem{f_{\bm{n}}}$ satisfies the 3D Helmholtz equation with the effective wavenumber $k_*$
\begin{equation}
    \label{eq:uniform_waves_expansion}
    \nabla^{2} \langle f_{\bm{n}} \rangle (\bm{r}_1) + k_*^2 \langle f_{\bm{n}} \rangle (\bm{r}_1) = 0 \quad \text{for } \quad \bm{r}_1 \in \mathcal R_a.
\end{equation}
Then, by using plane wave symmetry, shown in Appendix \ref{app:translation} with \eqref{eq:phase_dependency}, we conclude that
\begin{equation}
    \label{eq:plane_wave_basis}
    \langle f_{\bm{n}} \rangle (\bm{r}_1) = F_{\bm{n}} \mathrm e^{\mathrm i \bm{k}_* \cdot \bm{r}_1},
\end{equation}
where $\bm{k}_* = (k_{*x}, k_{*y}, k_{*z}) := \left( k_x, k_y, \sqrt{k_*^2 - k_x^2 - k_z^2} \right)$. We also choose $\mathrm{Im}[k_{*z}] > 0$, which is a form the Sommerfeld radiation condition and guarantees that we do not have unphysical waves.


Now, we take the limit $\delta \to 0$ to represent the case of a halfspace $\mathcal R$ filled with particles, as represented in \Cref{fig:halfspace}b. We also define the bulk region as done in \cite{Gower_2023}:
\begin{equation}
    \label{def:bulk_region}
    \mathcal R_{\text{Bulk}}:=\{ \bm{r}_1 \in \mathbb R^3 | z_1 > 2a \}.
\end{equation}

Below we follow the steps shown in \cite{Gower_2023} to deduce an effective wave equation and ensemble boundary conditions. we need to redo the steps here, as having a different background medium for the matrix does lead to some important differences. 

To start we note that 
\begin{multline}
    \label{eq:green_trick_1}
    (k_0^2 - k_*^2) \mathcal U_{\bm{n}'\bm{n}} (k \bm{r}_1 - k \bm{r}_2) \langle f_{\bm{n}'} \rangle(\bm{r}_2) = 
\\
    = \mathcal U_{\bm{n}'\bm{n}}(k_0 \bm{r}_1 - k_0 \bm{r}_2) \nabla_{\bm{r}_2}^2 \langle f_{\bm{n}'} \rangle(\bm{r}_2) - \langle f_{\bm{n}'} \rangle(\bm{r}_2) \nabla_{\bm{r}_2}^2 \mathcal U_{\bm{n}'\bm{n}}(k_0 \bm{r}_1 - k_0 \bm{r}_2),
\end{multline}
for $\bm{r}_1 \in \mathcal R_{\text{Bulk}}$ and $\bm{r}_2 \in \mathcal R_a$, because $\mathcal U_{\bm{n}'\bm{n}} (k \bm{r}_1 - k \bm{r}_2)$ and $ \langle f_{\bm{n}'} \rangle(\bm{r}_2)$ satisfy Helmholtz equations with wavenumbers $k_0$ and $k_*$ respectively. Then, by integrating the right side of \eqref{eq:green_trick_1} over $\bm{r}_2 \in \mathcal R_a \backslash \mathcal B(\bm{r}_1, 2a)$ and using Green's second identity, we get
\begin{equation}
    \label{eq:green_trick_2}
    \int_{\mathcal R_a \backslash \mathcal B(\bm{r}_1, 2a)} \mathcal U_{\bm{n}'\bm{n}} (k_0 \bm{r}_1 - k_0 \bm{r}_2) \langle f_{\bm{n}'} \rangle(\bm{r}_2) \mathrm d \bm{r}_2 \mathrm d \lambda = \frac{\mathcal I_{\bm{n}'\bm{n}}(\bm{r}_1) - \mathcal J_{\bm{n}'\bm{n}}(\bm{r}_1)}{(k_0^2 - k_*^2)},
\end{equation}
where $\mathcal I_{\bm{n}`\bm{n}}(\bm{r})$ and $\mathcal J_{\bm{n}`\bm{n}}(\bm{r})$ are given by
\begin{equation}
    \notag
    \begin{aligned}
        & \mathcal I_{\bm{n}'\bm{n}}(\bm{r}_1) := - F_{\bm{n}'}  \int_{\partial \mathcal R_a} \mathcal U_{\bm{n}'\bm{n}} (k_0 \bm{r}_1 - k_0 \bm{r}_2) \frac{\partial \mathrm e^{\mathrm i \bm{k}_* \cdot \bm{r}_2}}{\partial z_2} - \frac{\partial \mathcal U_{\bm{n}'\bm{n}} (k_0 \bm{r}_1 - k_0 \bm{r}_2)}{\partial z_2} \mathrm e^{\mathrm i \bm{k}_* \cdot \bm{r}_2} \mathrm d S_2,
\\
        & \mathcal J_{\bm{n}'\bm{n}}(\bm{r}_1) := F_{\bm{n}'} \int_{\partial \mathcal B(\bm{0}, 2a)} \mathcal U_{\bm{n}'\bm{n}} ( - k_0 \bm{r}') \frac{\partial \mathrm e^{\mathrm i \bm{k}_* \cdot (\bm{r}' + \bm{r}_1)}}{\partial z} - \frac{\partial \mathcal U_{\bm{n}'\bm{n}} (- k_0 \bm{r}')}{\partial z} \mathrm e^{\mathrm i \bm{k}_* \cdot (\bm{r}' + \bm{r}_1)} \mathrm d S',
    \end{aligned}
\end{equation}
with $\mathrm d S_2$ and $\mathrm d S'$ being the area elements. Also, in the second line of the above, we have performed the change of coordinates $\bm{r}_2 \to \bm{r}' + \bm{r}_1$.

We substitute expression \eqref{eq:green_trick_2} into \eqref{eq:final_gov_eq} to reach
\begin{equation}
    \label{eq:simplified_gov_eq}
        \begin{aligned}
        \langle f_{\bm{n}} \rangle (\bm{r}_1) = T_{\bm{n}} W_{\bm{n}} (\bm{r}_1) (\gamma_0 \zeta_T G - \zeta_R \langle B \rangle) + \mathfrak n \frac{T_{\bm{n}}}{k_0^2 - k_*^2}  \sum_{\bm{n}'} \left( \mathcal I_{\bm{n}'\bm{n}}(\bm{r}_1) - \mathcal J_{\bm{n}'\bm{n}}(\bm{r}_1) \right),
    \end{aligned}
\end{equation}
and we notice that $\langle f_{\bm{n}} \rangle(\bm{r}_1,\lambda)$ and 
$\mathcal J_{\bm{n}'\bm{n}}(\bm{r}_1)$ satisfy the Helmholtz equation with wavenumber $k_*$, while $W_{\bm{n}}(\bm{r}_1)$ and $\mathcal I_{\bm{n}'\bm{n}} (\bm{r}_1)$ satisfy the Helmholtz equation with wavenumber $k_0$. Because $k_0 \neq k_*$, see \cite[Appendix C]{gower2021effective}, we can split \eqref{eq:simplified_gov_eq} into
\begin{align}
    \label{eq:ensemble_wave_equation}
    &\langle f_{\bm{n}} \rangle (\bm{r}_1) +  \mathfrak n \sum_{\bm{n}'}\frac{T_{\bm{n}}}{k_0^2 - k_*^2} \mathcal J_{\bm{n}'\bm{n}}(\bm{r}_1) = 0,
\\
    \label{eq:ensemble_boundary_conditions}
    &W_{\bm{n}} (\bm{r}_1) \left( \gamma_0 \zeta_T G - \zeta_R \langle B \rangle \right) + \frac{\mathfrak n}{k_0^2 - k_*^2} \sum_{\bm{n}'} \mathcal I_{\bm{n}'\bm{n}}(\bm{r}_1) = 0.
\end{align}

Equation \eqref{eq:ensemble_wave_equation} is called the ensemble wave equation, and it is identical to \cite[eq. (4.7)]{gower2021effective}, which is expected because \eqref{eq:ensemble_wave_equation} is fully determined by the microstructure of the particulate material and not the exterior medium.

Equation \eqref{eq:ensemble_boundary_conditions} is similar to the ensemble boundary conditions in \cite[eq. 4.8]{gower2021effective}. However, \eqref{eq:ensemble_boundary_conditions} has one extra term representing the interaction between particles and the interface of the halfspace at $z = 0$.

Following the same steps as in \cite{gower2021effective}, one can use \eqref{eq:ensemble_wave_equation} to write down the following eigensystem for the eigenpair $(k_*, F_{\bm{n}})$:
\begin{equation}
    \label{eq:eigensystem}
    \tcboxmath{
        F_{\bm{n}} + 8 \pi a T_{\bm{n}}  \sum_{\bm{n}', \bm{n}_1} \frac{c_{\bm{n}' \bm{n} \bm{n}_1}}{k_*^2 - k_0^2} \mathrm i^{-\ell_1} \mathrm Y_{\bm{n}_1} (\hat{\bm{k}}_*) \mathrm N_{\ell_1} (2k_0a, 2k_*a) \mathfrak{n} F_{\bm{n}'}  = 0,
    }
\end{equation}
with $\bm{n}_1 = (\ell_1,m_1)$. The expression for $\mathrm N_{\ell}(x,z)$ and $c_{\bm{n} \bm{n}' \bm{n}_1}$ are defined in \cite[eqs. (5.5) and (B.4) respectively]{gower2021effective}.

\subsection{Normalisation condition}
\label{subsec:norm_cond}

The solution of \eqref{eq:eigensystem} provides the effective wavenumber $k_*$, however the eigenvectors $F_{\bm{n}}$ are determined only up to a multiplicative factor. The ensemble boundary condition \eqref{eq:ensemble_boundary_conditions} is needed to find a normalisation condition for $F_{\bm{n}}$, and fully determine the average field amplitude. To do so, we start by substituting \eqref{def:interface_coupling} and \cite[Eq. (7.10)]{gower2021effective} into \eqref{eq:ensemble_boundary_conditions}, leading to
\begin{equation}
    \label{eq:norm_cond_intermediate_step}
    \mathrm C_{\bm{n}} \mathrm e^{\mathrm i \bm{k}_0 \cdot \bm{r}_1} (\gamma_0 \zeta_T G - \zeta_R \langle B \rangle) = \mathrm C_{\bm{n}} \mathrm e^{\mathrm i \bm{k}_0 \cdot \bm{r}_1} \frac{\mathfrak n}{k_0^2 -  k_*^2} \sum_{\bm{n}'} \mathrm K_{\bm{n}'}(a) F_{\bm{n}'},
\end{equation}
where we defined 
\begin{equation}
    \notag
    \mathrm K_{(\ell', m')} (a) := \frac{2 \pi \mathrm i}{k_0 k_{0z}} (-\mathrm i)^{\ell'} \mathrm Y_{(\ell',m')}(\hat{\bm{k}}_0) (k_{*z} + k_{0z}) \mathrm e^{\mathrm i (k_{*z} - k_{0z}) a}.
\end{equation}

Then, we substitute \eqref{eq:plane_wave_basis} into \eqref{eq:average_scattered_coeff_B} to write down
\begin{equation}
    \label{eq:average_scattered_coeff_B_simplified}
    \langle B \rangle = - \frac{2\pi \mathfrak n}{k_0 k_{0z}}  \frac{\mathrm e^{\mathrm i (k_{*z} + k_{0z})a}}{\mathrm i (k_{*z} + k_{0z})} \sum_{(\ell,m)} \mathrm i^{\ell} \mathrm Y_{(\ell,m)} (\hat{\bm{k}}_0) F_{(\ell,m)},
\end{equation}
and we remind the reader we have already taken the limit $\delta \to 0$ at the beginning of \Cref{sec:effective_waves}.

Then, we substitute \eqref{eq:average_scattered_coeff_B_simplified} into \eqref{eq:norm_cond_intermediate_step} and divide both sides by $\mathrm C_{\bm{n}} \mathrm e^{\mathrm i \bm{k}_0 \cdot \bm{r}_1}$ to obtain
\begin{equation}
    \notag
    \gamma_0 \zeta_T G + \zeta_R \frac{2\pi \mathfrak n}{k_0 k_{0z}}  \frac{\mathrm e^{\mathrm i (k_{*z} + k_{0z})a}}{\mathrm i (k_{*z} + k_{0z})} \sum_{\bm{n}'} \mathrm i^{\ell'} \mathrm Y_{\bm{n}'} (\hat{\bm{k}}_0) F_{\bm{n}'} = \frac{\mathfrak n}{k_0^2 -  k_*^2} \sum_{\bm{n}'} \mathrm K_{\bm{n}'}(a) F_{\bm{n}'},
\end{equation}
where $\bm{n}' = (\ell', m')$.

Finally, we gather the terms which contain $F_{\bm{n}'}$ on the left-hand side to write down the following normalisation condition:
\begin{equation}
    \label{eq:norm_condition}
    \tcboxmath{
        \sum_{\bm{n}'} \mathrm M_{\bm{n}'} F_{\bm{n}'} = \gamma_0 \zeta_T G,
    }
\end{equation}
where
\begin{equation}
    \notag
    \mathrm M_{(\ell', m')} := \frac{\mathfrak n}{k_0^2 - k_*^2} \mathrm K_{(\ell', m')}(a) - \zeta_R \frac{2\pi \mathfrak n}{k_0 k_{0z}} \frac{\mathrm e^{\mathrm i (k_{*z} + k_{0z})a}}{(k_{*z} + k_{0z})} \mathrm i^{\ell'-1} \mathrm Y_{(\ell', m')} (\hat{\bm{k}}_0).
\end{equation}
This normalisation condition is the last equation needed to numerically calculate the eigenvectors $F_{\bm{n}}$, and together with $k_*$, they determine all the average amplitudes $\ensem{B}$, $\ensem{R}$ and $\ensem{A}$ through equations \eqref{eq:average_scattered_coeff_B} and \eqref{eq:sol_bound_cond_hs} respectively.

\subsection{High frequency limit}
\label{subsec:high freq limit}

As a side quest, here we deduce that, in the high frequency limit, the average reflection coefficient $\ensem{R}$ does not depend on the particle properties. In fact, $\ensem{R}$ is just the reflection from the matrix itself without particles in this limit. Results for high frequency, such as this, are not very common or well understood in the literature. 

We start by defining the volume fraction of particles as
\begin{equation}
    \label{def:vol_frac}
    \phi := \frac{|\mathcal P|}{|\mathcal R_a|} = \frac{4 \pi a^3 }{3} \mathfrak{n}.
\end{equation}
Substituting \eqref{def:vol_frac} into \eqref{eq:average_scattered_coeff_B_simplified}, one can deduce that
\begin{equation}
    \label{eq:abs_value_B}
    \begin{aligned}
        |\ensem{B}| =& \frac{3}{2} \frac{\phi \, \mathrm e^{-\mathrm{Im}[k_{*z}]a}}{ (k_0 a) (k_{0z} a) |k_{*z} + k_{0z} |a} \left| \sum_{(\ell,m)} \mathrm i^{\ell} \mathrm Y_{(\ell,m)} (\hat{\bm{k}}_0) F_{(\ell,m)} \right|
\\
        \leq& \frac{3}{2} \frac{\phi \, \mathrm e^{-\mathrm{Im}[k_{*z}]a}}{ (k_0 a) (k_{0z} a) |k_{*z} + k_{0z} |a} \sum_{\bm{n}} \left| \mathrm Y_{\bm n} (\hat{\bm{k}}_0) \right| \left| F_{\bm{n}} \right|,
    \end{aligned}
\end{equation}
which we will use to show that $|\ensem{B}| \to 0$ when $ka \to \infty$. 

Next, we write down the following relations between wavenumbers
\begin{equation}
    \notag
    k_0 = \frac{c}{c_0} k, \ k_{0z} = \sqrt{ \left( \frac{c}{c_0} \right)^2 k^2 - k_x^2 - k_y^2},
\end{equation}
from which we conclude that $k_0 a \to \infty$ and $k_{0z}a \to \infty$ when $ka \to \infty$ for a fixed angle of incidence. Using these limits in \eqref{eq:abs_value_B}, we conclude that $\ensem{B} \to 0$ with one added assumption: the absolute value of the eigenvectors $|F_{\bm{n}}|$ does not increase indefinitely for higher frequencies. This is reasonable because the norm of $F_{\bm{n}}$ is linked to the amplitude of the incident wave through the boundary condition \eqref{eq:ensemble_boundary_conditions}, though we have not been able to demonstrate this formally.

In other words, the response from the particles averages to zero due to incoherence for high frequency, and the reflected wave see only an empty halfspace with just the background matrix:
\begin{equation}
    \notag
    \ensem{R} \to \zeta_R G.
\end{equation}
It is possible to use this result to help calibrate an experimental measurement. As mentioned in the previous section, if we wish to use $\ensem{B}$ to characterise the particles, we need to subtract $ \zeta_R G$ from the average reflection coefficient $\ensem{R}$. If the background matrix properties are not known, one could perform scattering experiments while increasing the frequency until the reflected wave response stops changing. At this point, the reflection coefficient would equal $\zeta_R G$.

%
%
%

%

\subsection{Numerical results}
\label{subsec:num}

In this section we present some examples of numerical computations of the eigensystem \eqref{eq:eigensystem} and the normalisation condition \eqref{eq:norm_condition}, using the Julia library \cite{EffectiveWaves.jl}. The main purpose of these numerical results is to show that the system is easily solvable for broad frequencies and volume fractions. We also demonstrate how sensitive reflection is to average particle size, which could be useful for designing characterisation experiments. All infinite summations over double indices defined in \Cref{app:add_trans} were truncated at $\ell = 2$ for numerical computations.

These calculations performed in Julia provide the effective wavenumber $k_*$ and the eigenvectors $F_{\bm{n}}$. We substitute the expressions for  $k_*$ and $F_{\bm{n}}$ into \eqref{eq:average_scattered_coeff_B_simplified} to calculate the average backscattered amplitude $\ensem{B}$, and then substitute $\ensem{B}$ into \eqref{eq:sol_bound_cond_hs}$_{2}$ to calculate the average reflection coefficient $\ensem{R}$. 
To show how general the model is, and also provide some insight into the physics of the problem, we present three examples below with very different material properties.

\begin{figure}[ht!]
    \centering
    \begin{subfigure}[b]{0.48\textwidth}
        \centering
        \includegraphics[width=0.97\textwidth]{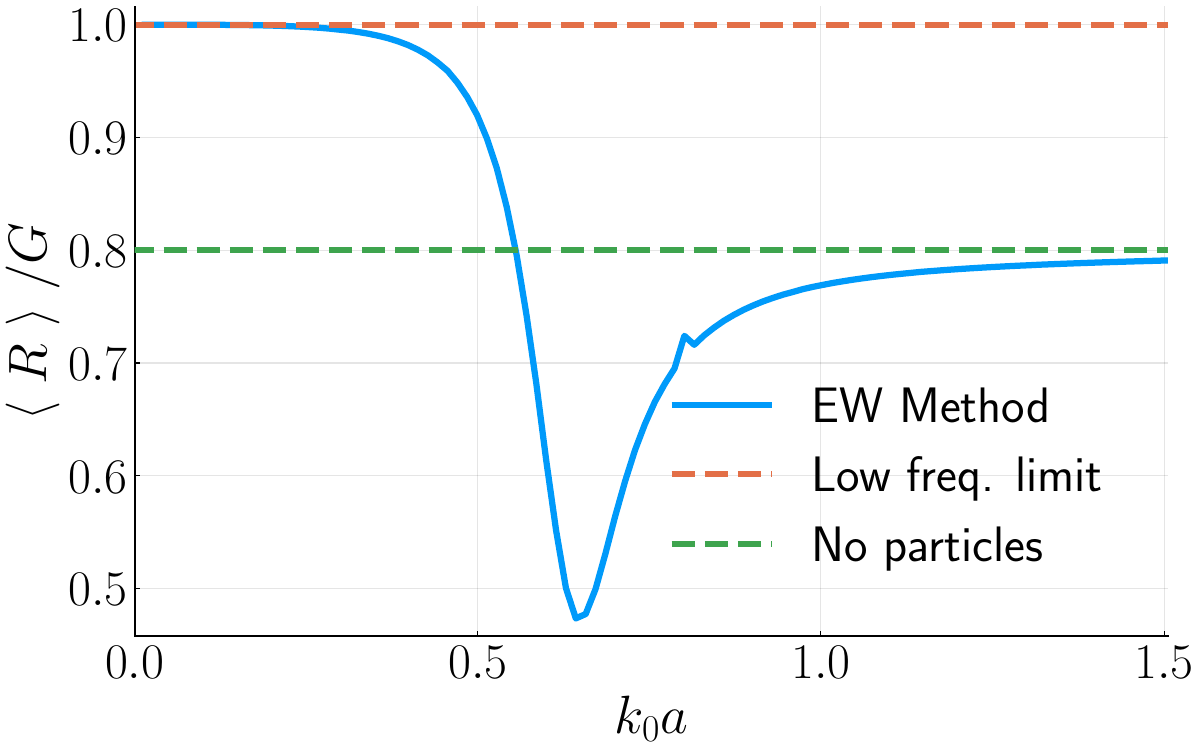}
        \caption{Normal incidence}
        \label{subfig:gas_normal}
    \end{subfigure}
    ~
    \begin{subfigure}[b]{0.48\textwidth}
        \centering
        \includegraphics[width=0.97\textwidth]{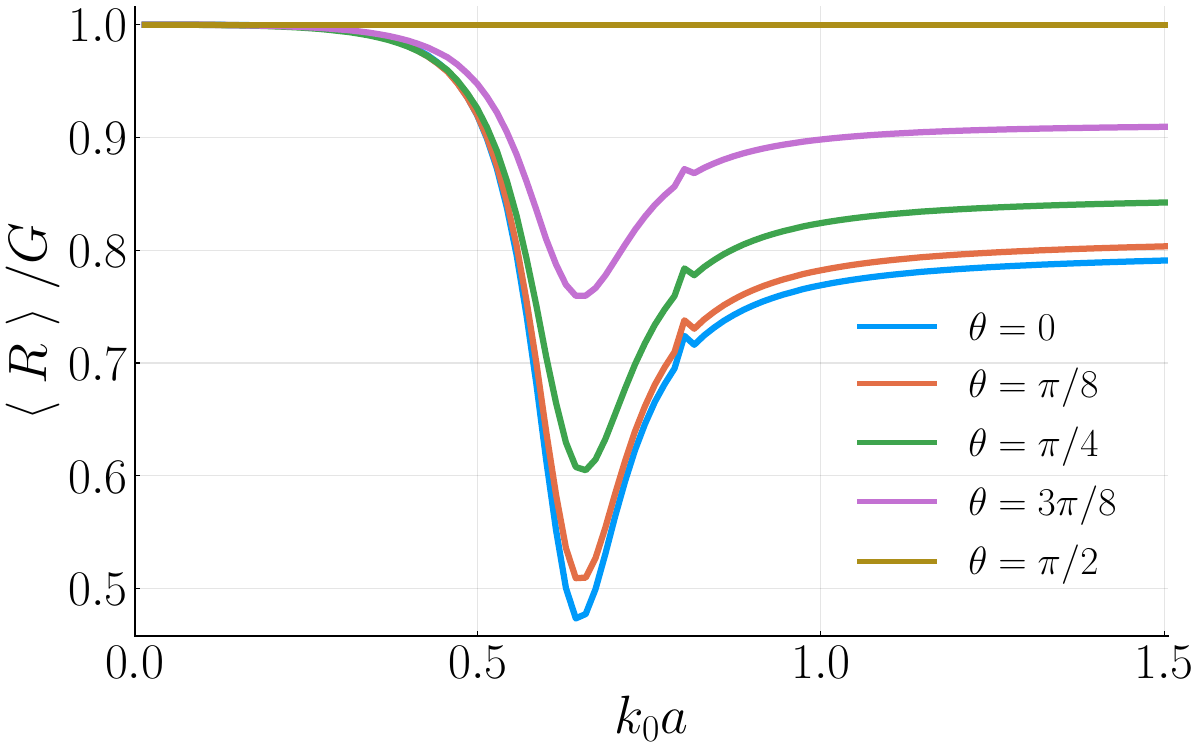}
        \caption{Varying angle of incidence}
        \label{subfig:gas_angle}
    \end{subfigure}
    \caption{Numerical computations of the average reflection coefficient $\ensem{R}$ divided by the incident wave amplitude $G$ by using the Effective waves method (``EW Method''). The particles are bubbles of gas in water ($c_s = c_0/100$, $\varrho_s = \varrho_0 / 100$) with volume fraction $\phi = 10\%$. The horizontal axis is the dimensionless frequency $k_0a$ and $\theta$ is the angle of incidence. On the left, the plane wave incidence is normal ($k_x = k_y = 0$), while on the right the angle of incidence is varying ($k_x \geq 0$, $k_y = 0$). The dashed lines represent the low frequency limit response and the reflection without any particles in the matrix. The exterior medium is a solid ($c = 3c_0$, $\varrho = 3\varrho_0$).}
    \label{fig:gas_bubbles}
\end{figure}

\textbf{Case 1.} We consider that the matrix is water, the exterior medium is a solid with acoustic properties $c = 3 c_0$ and $\varrho = 3 \varrho_0$, and particles some gas with $c_s = c_0/100$ and $\varrho_s = \varrho_0/100$. The reflection coefficients $\ensem{R}$ are presented in \Cref{fig:gas_bubbles} below for the case of volume fraction $\phi = 10\%$.

In \Cref{fig:gas_bubbles}, the low frequency limit is computed with the formulas provided in \cite{gower2021effective}, and the reflection coefficient 
 of a homogeneous halfspace with no particles is given by $R = \zeta_R G$. The agreement between low frequency limit and the Effective waves method for $ka \ll 1$ is a good sanity check for the formulas for the normalisation condition \eqref{eq:norm_condition}.

Three main results can be obtained from \Cref{fig:gas_bubbles}: 1) The average reflection is sensitive to particle radius, with a drop of more than 50\% in \Cref{subfig:gas_normal} if varying radius $a$ for a fixed frequency $\omega$; 2) we have numerical evidence that the high frequency limit matches the case with no particles; and 3) changing the angle of incidence in \Cref{subfig:gas_angle} only makes reflection less sensitive to particle radius. The last result suggests that normal incidence should be the optimal strategy to sense particle size if $\ensem{R}$ can be measured with only one angle of incidence.

\textbf{Case 2.} We consider that the matrix is a solid medium ($c_0 = 3 c$ and $\varrho_0 = \varrho$), and the exterior medium is water. In this case, we choose solid inclusions in the matrix such that $c_s = 10c$ and $\varrho_s = 10\varrho$. The results are presented in \Cref{fig:harder} below.
\begin{figure}[hb!]
    \centering
    \begin{subfigure}[b]{0.48\textwidth}
        \centering
        \includegraphics[width=0.97\textwidth]{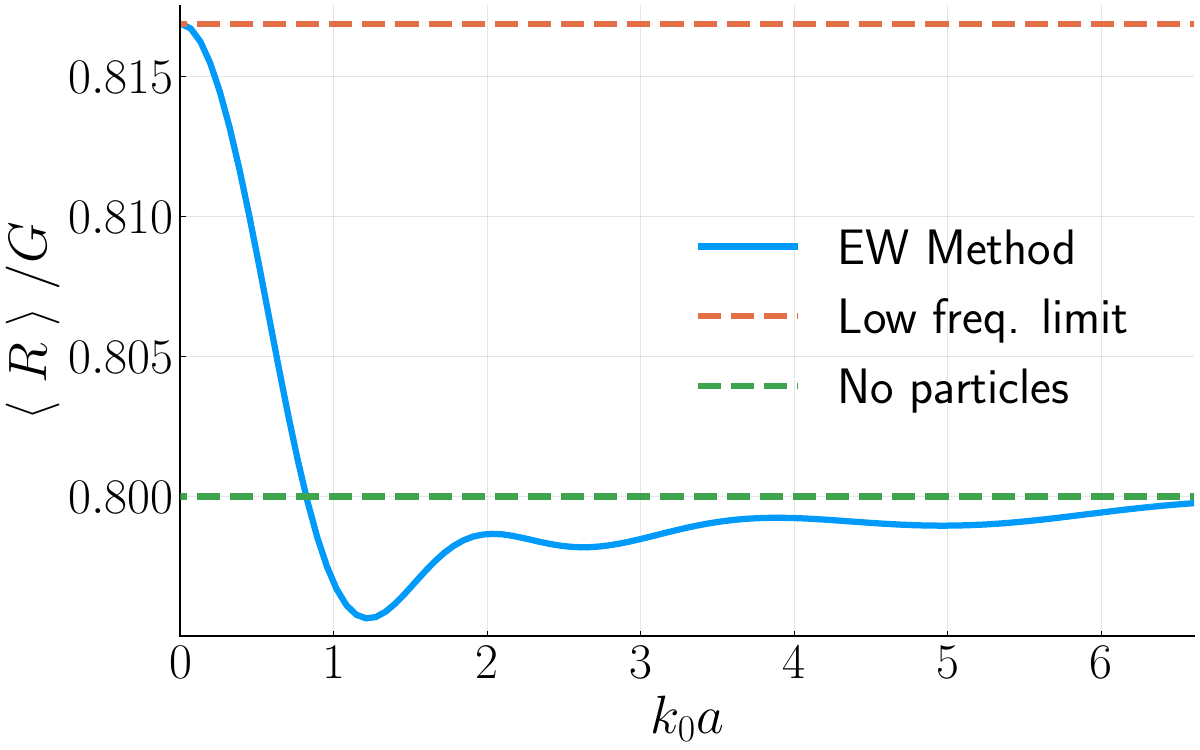}
        \caption{Normal incidence}
        \label{subfig:harder_normal}
    \end{subfigure}
    ~
    \begin{subfigure}[b]{0.48\textwidth}
        \centering
        \includegraphics[width=0.97\textwidth]{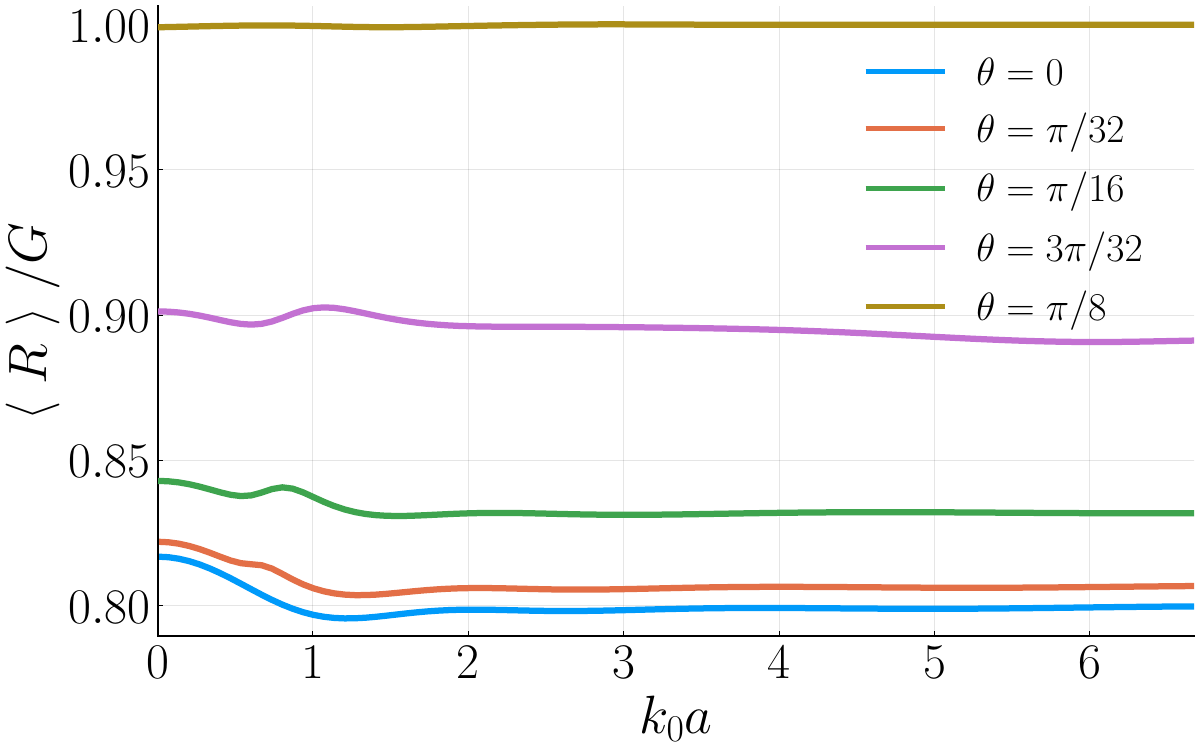}
        \caption{Varying angle of incidence}
        \label{subfig:harder_angle}
    \end{subfigure}
    \caption{Numerical computations for dimensionless average reflection using the Effective waves method (``EW Method''). The exterior medium is water, and the matrix is solid ($c_0 = 3c$, $\varrho_0 = 3\varrho$). The particles are solid inclusions ($c_s = 10c$, $\varrho_s = 10\varrho$) with volume fraction $\phi = 10\%$. The horizontal axis is the dimensionless frequency $k_0a$ and $\theta$ is the angle of incidence. On the left, the plane wave incidence is normal ($k_x = k_y = 0$), while on the right the angle of incidence is varying ($k_x \geq 0$, $k_y = 0$). The dashed lines represent the low frequency limit response and the reflection without any particles in the matrix.}
    \label{fig:harder}
\end{figure}

In this case 2, \Cref{fig:harder} shows the same qualitative behaviour as in \Cref{fig:gas_bubbles}. However, two observations must be done: 1) the average reflection is less sensitive to particle radius, with only about a drop of about 3$\%$ in \Cref{subfig:harder_normal} when varying radius $a$ for a fixed frequency $\omega$; and 2) in \Cref{subfig:harder_angle}, total reflection happens for angle of incidence $\theta$ bigger than the critical angle for the homogeneous matrix without particles, given by
\begin{equation}
    \notag
    \theta_c = \arcsin \left({\frac{c}{c_0}}\right),
\end{equation}
which for case 2 is equal to $\theta_c \approx 0.34 \textrm{rad} < \pi / 8$.

\textbf{Case 3.} To study how reflection changes when varying the volume fraction of particles \eqref{def:vol_frac}, we compute the average reflection coefficient for hard solid particles in water, $c_s = 100c$ and $\varrho_s = 100\varrho_0$. We take the exterior medium as a softer solid ($c = 3 c_0$ and $\varrho = \varrho_0$). The results are presented in \Cref{fig:powder_vol_frac} below.
\begin{figure}[ht!]
    \centering
    \includegraphics[width=0.54\textwidth]{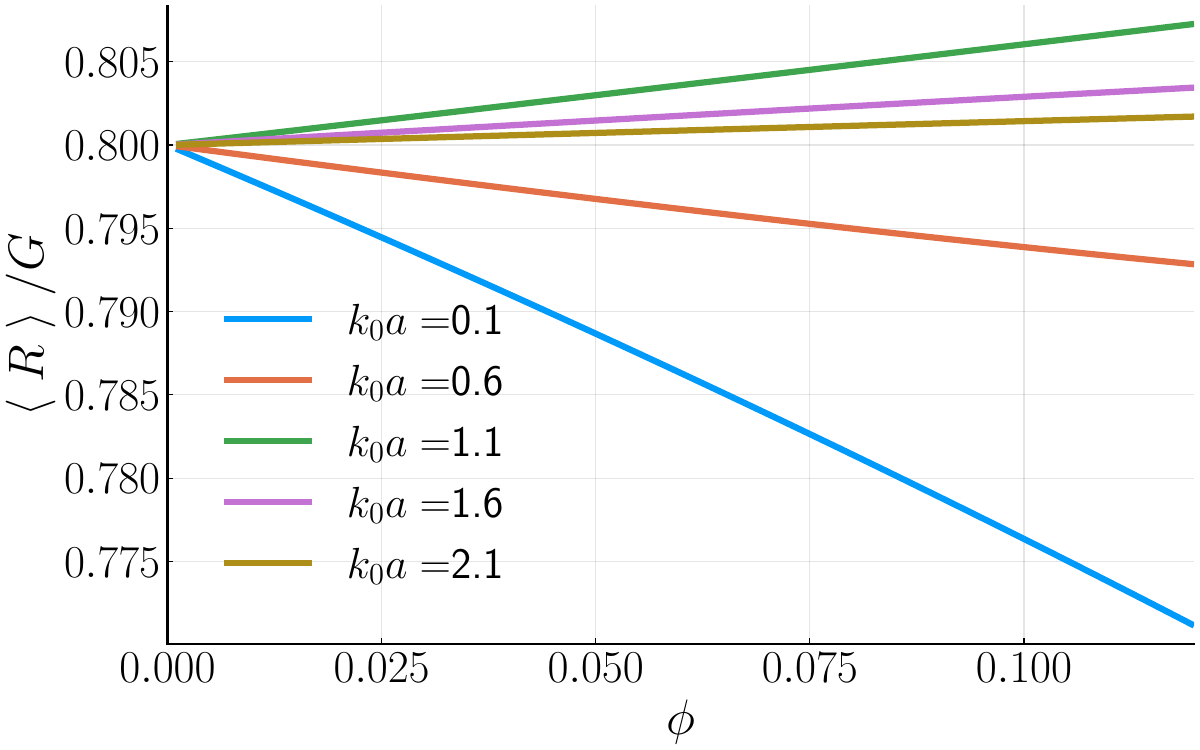}
    \caption{Numerical computations for dimensionless average reflection using the Effective waves method (``EW Method''). We have a hard powder in water ($c_s = 100c_0$, $\varrho_s = 100\varrho$) for some values of dimensionless frequent $k_0 a$. The horizontal axis is the volume fraction $\phi$. The plane wave incidence is normal ($k_x = k_y = 0$) and the exterior medium is a solid ($c = 3c_0$, $\varrho = 3\varrho_0$).}
    \label{fig:powder_vol_frac}
\end{figure}

\Cref{fig:powder_vol_frac} shows that the dependency of $\ensem{R}$ with respect to volume fraction can be approximately a simple linear relation. That suggests a first-order expansion in volume fraction would be accurate for reflection measurements for case 3 depicted in \Cref{fig:powder_vol_frac}. We also notice that the slope of the curves in \Cref{fig:powder_vol_frac} is sensitive to radius $a$, which may be useful for sensing methods for the particle size distribution.

\section*{Conclusions}
\label{sec:conc}

In this work, we solve an open challenge on how to calculate sound wave scattering from particles embedded in a matrix with different acoustic properties than the exterior medium. It seems we are the first to reach simple solvable equations, while retaining the same order of approximation used in other methods for dense particles, the Quasi-Crystalline Approximation (QCA) \cite{Lax_QCA_1952}. The QCA is one of the most successful approximations that captures multiple scattering between particles \cite{Varadans_84,Linton_2005, gower2019multiple}.

The method we develop can be applied to any material geometry, but for simplicity, we explore the simplest case in this paper: 
plane wave incidence on a halfspace. 


\textbf{Theoretical results.} The key to solve for more than one background medium was to reformulate the QCA in terms of an approximation involving the field that excites each particle \eqref{eq:exciting_field_average}.
This reformulation shows that QCA is equivalent to averaging over self-interactions for all orders of multiple scattering. A similar interpretation has been noted before by \cite{Mishchenko_2011} by comparing QCA with a diagrammatic approximation given by Twersky \cite{twersky1964}. However, unlike the Twersky approximation, QCA does not discard any scattering contributions if defined as \eqref{eq:xqca-approximation}$_{(1)}$ or \eqref{eq:exciting_field_average}. Instead, QCA averages over self-interactions.
Our reformulation confirm that, for a big class of ensembles presented in \Cref{sec:ensemble}, QCA is exact up to second-order scattering, as demonstrated in \cite{Martin2008}.

Our reformulation of QCA enabled us to reach simple solvable equations for the case of multiple background mediums, we call the extension of QCA to multiple mediums the eXtended Quasi-Crystalline Approximation (X-QCA). Using X-QCA leads to significant simplification in comparison to method that use only QCA \cite{Slab_book}, and is applicable for a broad range of frequencies and particle properties.

\textbf{Numerical results.} To exemplify how the method works for a broad range of frequencies, volume fractions and material properties, we present explicit computations of the average reflection coefficient $\ensem{R}$ in \Cref{subsec:num}. We show numerically that the average reflection coefficient can be sensitive to particle radius, depending on the material properties. This result can contribute to the development of sensing methods for particle size distribution of powders, emulsions, porous material or slurries.

We also give numerical evidence that, in the high frequency limit, there is no coherent reflection due to particles. This means the high frequency response can be used as calibration for equipment that uses the reflection coefficient to take measurements of particle scattering. If such measurements are not available, the simple measurement of reflected waves from the matrix with no particles can also be used as calibration, see discussion in \Cref{sec:final_average_gov_eqs}. 

\textbf{Possible generalisations.} The model presented makes use of the Effective Waves Method \cite{gower2021effective}, which allows any geometry or acoustic properties. The same procedure in this work can be used to compute the average wave scattering by a compact material filled with particles. On top of that, our method can be easily applied to any linear waves, including electromagnetic and elastic waves. 

\textbf{Future work.} The method introduced here is not a mere theoretical curiosity. It is a necessary step before reaching quantifiable methods to measure particles in many areas of science and engineering. It is needed because, in almost any sensor setup, there will be different background mediums. To account for these mediums, and calculate the average reflection and transmission, one would need our results. A simple example is illustrated in \Cref{fig:pipe} which involves a particulate flowing through a pipe past a sensor. 
In other words, a clear future avenue is to use the results here to develop sensors to measure particulates from reflection and transmission.

Other future avenues include validating our method (X-QCA) against high fidelity Monte Carlo simulations or experimental data. Another interesting step is to apply our results to design layered media containing particulates, as these can form functionally graded materials \cite{Miyamoto_1999}, which have applications in many areas like aerospace engineering, nuclear energy, biomedical, optics and others \cite{Li_2018}. 

\section*{Acknowledgements}

The authors would like to acknowledge Aristeidis Karnezis, Brandon O'Connell and Gerhard Kristensson for their insightful discussions. Paulo Piva
gratefully acknowledges funding from an EPSRC Case studentship with Johnson Matthey. Kevish Napal and Art Gower gratefully acknowledge support from EPSRC (EP/V012436/1).

\section*{Appendices}

\appendix

\section{Addition translation matrices}
\label{app:add_trans}

Here we provide the notation for addition translation matrices for spherical Bessel waves. For a translation $\bm{d}$ ($\bm{y}=\bm{x}+\bm{d}$), we have
\begin{equation}
\label{def:spherical_trans_matrices}
    \begin{aligned}
        \mathrm v_{\bm{n}}(\bm{y})& = \sum_{\bm{n}'}\mathcal{V}_{\bm{nn}'}(\bm{d})\mathrm v_{\bm{n}'}(\bm{x}) \quad \forall \ \bm{x},\bm{d}\in \mathbb{R}^3,
\\
        \mathrm u_{\bm{n}}(\bm{y})& = \sum_{\bm{n}'}\mathcal{V}_{\bm{nn}'}(\bm{d})\mathrm u_{\bm{n}'}(\bm{x}) \quad \text{for } |\bm{x}| > |\bm{d}|,
\\
        \mathrm u_{\bm{n}}(\bm{y})& = \sum_{\bm{n}'}\mathcal{U}_{\bm{nn}'}(\bm{d})\mathrm v_{\bm{n}'}(\bm{x}) \quad \text{for } |\bm{x}| < |\bm{d}|,
    \end{aligned}
\end{equation}
where the summations over double indices $\bm{n} = (\ell, m)$ represent double sums defined as
\begin{equation}
    \notag
    \sum_{\bm{n}} := \sum_{\ell = 0}^{\infty} \sum_{m = -\ell}^{\ell}.
\end{equation}
The addition translation matrices $\mathcal{V}_{\bm{nn}'}(\bm{d})$ and $\mathcal{U}_{\bm{nn}'}(\bm{d})$ are given by \cite[eq. (B.3)]{gower2021effective}.

\section{Regular spherical to plane waves}
\label{app:v_to_plane}

In this section, we determine the expression that connects the spherical Bessel waves representation of the regular field \eqref{eq:average_reg_field_inside} with its planar wave representation \eqref{eq:reg_field_big}. These expressions are only valid after the $L \to \infty$ limit in \Cref{subsec:inf_J}. For that, we will use the following definition for the spherical harmonics
\begin{equation}
    \notag
    \mathrm Y_{(\ell, m)} (\hat{\bm{r}}) := (-1)^m \sqrt{\frac{2\ell + 1}{4 \pi}\frac{(\ell - m)!}{(\ell + m)!}} \mathrm P_{\ell}^m(\cos\theta) \mathrm e^{\mathrm i m \phi},
\end{equation}
where $\mathrm P_{\ell}^m$ are the usual associated Legendre polynomials.

We equate the averaged field expressed in both spherical Bessel \eqref{eq:average_reg_field_inside} and planar representations \eqref{eq:reg_field_big}, having in mind we chose $k_{0z}$ such that $\ensem{A_-} = 0$. This results in
\begin{equation}
    \notag
    \langle A \rangle \mathrm e^{\mathrm i \bm{k}_0 \cdot \bm{r}} = \sum_{\bm{n}} \ensem{g_{\bm{n}}} \mathrm v_{\bm{n}}(k_0\bm{r})
\end{equation}
Then, we substitute the planar wave expansion in terms of spherical Bessel waves, given by
\begin{equation}
    \label{eq:plane_to_v}
    \mathrm e^{\mathrm i \bm{k} \cdot \bm{r}} = 4 \pi \sum_{\ell=0}^{\infty} \sum_{m = -\ell}^{\ell} \mathrm i^{\ell}  \overline{\mathrm Y_{(\ell,m)}} (\hat{\bm{k}}) \mathrm v_{(\ell,m)}(k \bm r) \mathrm ,
\end{equation}
where the overline denotes complex conjugation. The final step is to use the orthogonality relations of the spherical harmonics, which leads to
\begin{equation}
    \label{eq:relation_C_g}
    \langle g_{\bm{n}} \rangle = \mathrm C_{\bm{n}} \langle A \rangle,
\end{equation}
where we have defined 
\begin{equation}
\label{def:C_coefs}
    \mathrm C_{(\ell, m)} := 4 \pi \mathrm i^{\ell} \, \overline{ \mathrm Y_{(\ell, m)}}(\hat{\bm{k}}_0).
\end{equation}

\section{Translation symmetry}
\label{app:translation}

In the limit of an infinite number of particles (see \Cref{subsec:inf_J}), the average regular field can be represented by a plane wave, given by \eqref{eq:reg_field_big}. Then, its dependency on $x$ and $y$ is given only by a known complex phase. For a general translation in the $z = 0$ plane of $\bm{b} = x_0 \hat{\bm{x}} + y_0 \hat{\bm{y}}$, we have
\begin{equation}
    \notag
    \begin{aligned}
        \langle u_{\text{reg}} (\bm{r} + \bm{b}) \rangle &= \langle u_{\text{reg}} (\bm{r}) \rangle \mathrm e^{\mathrm i \bm{k}_0 \cdot \bm{b}} = \sum_{\bm{n}'} \ensem{g_{\bm{n}'}}  \mathrm v_{\bm{n}'}(k_0 \bm{r}) \mathrm e^{\mathrm i \bm{k}_0 \cdot \bm{b}},
\\
        \langle u_{\text{reg}} (\bm{r} + \bm{b}) \rangle &= \sum_{\bm{n}} \ensem{g_{\bm{n}}} \mathrm v_{\bm{n}}(k_0 \bm{r} + k_0 \bm{b}) = \sum_{\bm{nn}'} \ensem{g_{\bm{n}}} \mathcal{V}_{\bm{nn}'}(k_0 \bm{b}) \mathrm v_{\bm{n}'}(k_0 \bm{r}),
    \end{aligned}
\end{equation}
where we have used \eqref{eq:average_reg_field_inside} in both equations above, and translation matrices in Appendix \ref{app:add_trans} in the second line. Equating both lines above, and using the orthogonality relations of spherical harmonics, we conclude that
\begin{equation}
    \label{eq:regulaR_aoefficient_symmetry}
    \ensem{g_{\bm{n}}}  \mathrm e^{\mathrm i \bm{k}_0 \cdot \bm{b}} = \sum_{\bm{n}'} \ensem{g_{\bm{n}'}} \mathcal{V}_{\bm{n}'\bm{n}}(k_0 \bm{b}),
\end{equation}

Then, we perform the same translation of $\bm{b}$ in \eqref{eq:gov_eq_inside_average_qca} to get 
\begin{equation}
    \begin{aligned}
    \notag
        \langle f_{\bm{n}} \rangle (\bm{r}_1 + \bm{b}) = & \  T_{\bm{n}} \sum_{\bm{n}'} \ensem{g_{\bm{n}'}} \mathcal{V}_{\bm{n}'\bm{n}} (k_0 \bm{r}_1 + \bm{b})+
\\
        & + \mathfrak n \, T_{\bm{n}} \sum_{\bm{n}'} \int_{\mathcal{R}_a \backslash \mathcal{B}(\bm{r}_1 + \bm{b}, 2a)} \mathcal{U}_{\bm{n}'\bm{n}}(k_0 \bm{r}_1 - k_0 \bm{r}_2 + k_0 \bm{b}) \langle f_{\bm{n}'} \rangle (\bm{r}_2) \mathrm d \bm{r}_2.
    \end{aligned}
\end{equation}
We decompose the regular translation matrix into two factors (see \cite[eq. (B.3)]{gower2021effective}) and perform the change of variables $\bm{r}_2 \to \bm{r}_2' = \bm{r}_2 - \bm{b}$ into the above to reach
\begin{equation}
    \notag
    \begin{aligned}
        \langle f_{\bm{n}} \rangle (\bm{r}_1 + \bm{b}) =& \  T_{\bm{n}} \sum_{\bm{n}' \bm{n}''} \ensem{g_{\bm{n}'}} \mathcal{V}_{\bm{n}' \bm{n}''} (k_0 \bm{b}) \mathcal{V}_{\bm{n}'' \bm{n}} (k_0 \bm{r}_1) +
\\
        & + \mathfrak n \, T_{\bm{n}} \sum_{\bm{n}'} \int_{\mathcal{R}_a \backslash \mathcal{B}(\bm{r}_1, 2a)} \mathcal{U}_{\bm{n}'\bm{n}}(k_0 \bm{r}_1 - k_0 \bm{r}_2') \langle f_{\bm{n}'} \rangle (\bm{r}_2' + \bm{b}) \mathrm d \bm{r}_2'.
    \end{aligned}
\end{equation}
Finally, we substitute \eqref{eq:regulaR_aoefficient_symmetry} in the above, leading to
\begin{equation}
    \notag
    \begin{aligned}
        \langle f_{\bm{n}} \rangle (\bm{r}_1 + \bm{b}, \lambda) =& \  T_{\bm{n}} \sum_{\bm{n}''} \ensem{g_{\bm{n}''}} \mathcal{V}_{\bm{n}'' \bm{n}} (k_0 \bm{r}_1) \mathrm e^{\mathrm i \bm{k}_0 \cdot \bm{b}} +
\\
        & + \mathfrak n \, T_{\bm{n}} \sum_{\bm{n}'} \int_{\mathcal{R}_a \backslash \mathcal{B}(\bm{r}_1, 2a)} \mathcal{U}_{\bm{n}'\bm{n}}(k_0 \bm{r}_1 - k_0 \bm{r}_2') \langle f_{\bm{n}'} \rangle (\bm{r}_2' + \bm{b}) \mathrm d \bm{r}_2',
    \end{aligned}
\end{equation}
which is the same equation as \eqref{eq:gov_eq_inside_average_qca}, but written in terms of $\langle f_{\bm{n}} \rangle (\bm{r}_1 + \bm{b}) = \langle f_{\bm{n}} \rangle (\bm{r}_1) \mathrm e^{\mathrm i \bm{k}_0 \cdot \bm{b}}$ instead of $\langle f_{\bm{n}} \rangle (\bm{r}_1)$. Assuming the uniqueness of the solution for \eqref{eq:gov_eq_inside_average_qca}, one can deduce that 
\begin{equation}
    \label{eq:phase_dependency}
    \langle f_{\bm{n}} \rangle (\bm{r}_1) = \langle f_{\bm{n}} \rangle (0,0,z_1) \mathrm e^{\mathrm i (k_x x_1 + k_y y_1)}, 
\end{equation}
and we have reduced the dimensions of our integral equation from three to one, due to symmetry.

\section{Outgoing spherical to plane waves}
\label{app:planar_to_u}

To simplify \eqref{eq:average_particle_scattering_inf_J}, we use \eqref{eq:phase_dependency} from \Cref{app:translation} above, which leads to
\begin{equation}
    \label{eq:planar_sym_simplification_scattered_waves}
    \lim_{\substack{J \to \infty \\ L \to \infty}} \sum_i \ensem{u_\text{sc}^i(\bm r)} = \mathfrak{n} \sum_{\bm{n}} \int_{a + \delta}^{\infty} \langle f_{\bm{n}} \rangle (0,0,z_1) I_{\bm{n}}(\bm r,z_1) \mathrm d z_1,
\end{equation}
where we have defined the following quantity that can be determined analytically:
\begin{equation}
    \label{eq:sph_hankel_integral}
    I_{\bm{n}}(\bm r,z_1):=\int_{\mathbb R^2} \mathrm u_{\bm{n}}(k_0\bm{r}-k_0\bm{r}_1) \mathrm e^{\mathrm i (k_x x_1 + k_y y_1)} \mathrm d x_1 \mathrm d y_1, \quad z \neq z_1,
\end{equation}
and results in a plane wave in the $z_1$ direction. To simplify \eqref{eq:sph_hankel_integral}, we use the following transformation formula \cite{10.1063/1.1704333,Bostrom+Kristensson+Strom1991,Kristensson2016,gerhard_thesis,gower2021effective}
\begin{equation}
    \notag
    \mathrm u_{\bm{n}} (k_0 \bm{r}) = \mathrm u_{(\ell, m)} (k_0 \bm{r}) = \frac{1}{2 \pi \mathrm i^\ell} \int_{\mathbb R^2} \frac{\mathrm Y_{\bm n} (\hat{\bm{q}})}{k_0 q_z} \mathrm e^{\mathrm i \bm{q} \cdot \bm{r}} \mathrm d q_x \mathrm d q_y, \quad z > 0
\end{equation}
where $\bm{q} = (q_x, q_y, q_z)$ with $q_x^2 + q_y^2 + q_z^2 = k_0^2$.  We substitute the above into \eqref{eq:sph_hankel_integral} and perform the following calculations for $z > z_1$ as follows:
\begin{equation}
    \label{eq:spherical_integration_1}
    \begin{aligned}
        I_{\bm{n}}(\bm r,z_1) &= \frac{1}{2 \pi \mathrm i^\ell} \int_{\mathbb R^2} \left[
        \int_{\mathbb R^2} \frac{\mathrm Y_{\bm n} (\hat{\bm{q}})}{k_0 q_z} e^{\mathrm i \bm{q} \cdot (\bm{r} - \bm{r}_1) + \mathrm i (k_x x_1 + k_y y_1)} \mathrm d q_x \mathrm d q_y \right] \mathrm d x_1 \mathrm d y_1
\\
        &= \frac{1}{2 \pi \mathrm i^\ell} \int_{\mathbb R^2} \frac{\mathrm Y_{\bm n} (\hat{\bm{q}})}{k_0 q_z} \mathrm e^{\mathrm i \bm{q} \cdot \bm{r} - \mathrm i q_z z_1}
        \left[ \int_{-\infty}^{\infty} e^{\mathrm i(k_x - q_x)x_1} \mathrm d x_1 \int_{-\infty}^{\infty} e^{\mathrm i(k_y - q_y)y_1} \mathrm d y_1 \right] \mathrm d q_x \mathrm d q_y
\\
        &= \frac{2\pi}{\mathrm i^\ell k_0} \int_{\mathbb R^2} \frac{\mathrm Y_{\bm n} (\hat{\bm{q}})}{q_z} \mathrm e^{\mathrm i (\bm{q} \cdot \bm{r} - q_z z_1)}
        \delta(q_x - k_x) \delta(q_y - k_y) \mathrm d q_x \mathrm d q_y
\\
        &= \frac{2\pi \mathrm i^{-\ell}}{k_0 k_{0z}} \mathrm Y_{\bm n} (\hat{\bm{k}}) \mathrm e^{\mathrm i (k_x x + k_y y) +\mathrm i k_{0z} (z - z_1)}
    \end{aligned}
\end{equation}
where we have changed the order of integration, and used the Fourier expansion of the Dirac delta distribution:
\begin{equation}
    \notag
    \delta(q) = \frac{1}{2 \pi}\int_{-\infty}^{\infty} \mathrm e^{- \mathrm i q x} \mathrm d x.
\end{equation}

For the case $z < z_1$, we use the fact that 
\begin{equation}
    \notag
    I_{\bm{n}} (\bm{r}, z_1) = (-1)^\ell \int_{\mathbb R^2} \mathrm u_{\bm{n}}(k_0\bm{r}_1-k_0\bm{r})  \mathrm e^{\mathrm i (k_x x_1 + k_y y_1)} \mathrm d x_1 \mathrm d y_1; \quad z \neq z_1,
\end{equation}
and we repeat the same computations in \eqref{eq:spherical_integration_1} to reach
\begin{equation}
    \label{eq:spherical_integration_2}
    I_{\bm{n}} (\bm{r}, z_1) = \frac{2\pi \mathrm i^\ell}{k_0 k_{0z}} \mathrm Y_{\bm n} (\hat{\bm{k}}_0) \mathrm e^{\mathrm i (k_x x + k_y y) +\mathrm i k_{0z} (z_1 - z)}.
\end{equation}

Substituting \eqref{eq:spherical_integration_2} into \eqref{eq:planar_sym_simplification_scattered_waves}, we conclude that the sum of the average scattered waves by particles \eqref{eq:average_particle_scattering} can also be represented as a plane wave in the region $0 < z < \delta$. The results of this appendix motivate the definition of the average backscattered amplitude $\ensem{B}$ in \eqref{eq:average_scattered_coeff_B}.

\printbibliography

@article{waterman1961multiple, 
  title={Multiple scattering of waves},
  author={Waterman, Peter Cary and Truell, Rohn},
  journal={Journal of mathematical physics},
  volume={2},
  number={4},
  pages={512--537},
  year={1961},
  publisher={American Institute of Physics}
}

@article{Willis2023,
  title={Transmission and reflection of energy at the boundary of a random two-component composite},
  author={J. R. Willis},
  journal={Proc. R. Soc. A},
  volume={479},
  number = {20220730},
  doi = {10.1098/rspa.2022.0730},
  year={2023}
}

@article{Willis2020,
  title={Transmission and reflection at the boundary of a random two-component composite},
  author={J. R. Willis},
  journal={Proc. R. Soc. A},
  volume={476},
  number = {20190811},
  doi = {10.1098/rspa.2019.0811},
  year={2020}
}

@article{Willis2019,
  title={Transmission and reflection of waves at an interface between ordinary material
and metamaterial},
  author={J. R. Willis},
  journal={J. Mech. Phys. Solids},
  volume={136},
  number = {103678},
  doi = {10.1016/j.jmps.2019.103678},
  year={2019}
}

@article{Willis1981,
title = {Variational principles for dynamic problems for inhomogeneous elastic media},
journal = {Wave Motion},
volume = {3},
number = {1},
pages = {1-11},
year = {1981},
doi = {https://doi.org/10.1016/0165-2125(81)90008-1},
author = {J.R. Willis}
}

@article{fikioris1964multiple,
  title={Multiple Scattering of Waves. II.``Hole Corrections''in the Scalar Case},
  author={Fikioris, JG and Waterman, PC},
  journal={Journal of Mathematical Physics},
  volume={5},
  number={10},
  pages={1413--1420},
  year={1964},
  publisher={American Institute of Physics}
}

@book{kong2004scattering,
  title={Scattering of electromagnetic waves: numerical simulations},
  author={Kong, Jin Au and Tsang, Leung and Ding, Kung-Hau and Ao, Chi On},
  year={2004},
  publisher={John Wiley \& Sons}
}

@article{fawcett2021effective,
  title={The effective medium for a cylinder with cylindrical inclusions},
  author={Fawcett, John A},
  journal={The Journal of the Acoustical Society of America},
  volume={150},
  number={4},
  pages={2600--2612},
  year={2021},
  publisher={Acoustical Society of America}
}

@article{MISHCHENKO1996,
title = {T-matrix computations of light scattering by nonspherical particles: A review},
journal = {Journal of Quantitative Spectroscopy and Radiative Transfer},
volume = {55},
number = {5},
pages = {535-575},
year = {1996},
note = {Light Scattering by Non-Spherical Particles},
issn = {0022-4073},
doi = {10.1016/0022-4073(96)00002-7},
author = {Michael I. Mishchenko and Larry D. Travis and Daniel W. Mackowski}
}

@article{gower_reflection_2018,
	title = {Reflection from a multi-species material and its transmitted effective wavenumber},
	volume = {474},
	issn = {1364-5021, 1471-2946},
	doi = {10.1098/rspa.2017.0864},
	language = {en},
	number = {2212},
	journal = {Proc. R. Soc. A},
	author = {Gower, Artur L. and Smith, M. J. A. and Parnell, W. J. and Abrahams, I. D.},
	year = {2018},
	pages = {20170864}
}

@article{2020MultipleScatering.jl,
  title={MultipleScatering.jl: A Julia library for simulating, processing, and plotting multiple scattering of waves.},
  author={Artur L Gower and J Deakin},
  journal={ \href{https://github.com/JuliaWaveScattering/MultipleScattering.jl}{Github}},
  year={2020}
}

@article{gower2019multiple,
  title={Multiple waves propagate in random particulate materials},
  author={Gower, Artur L. and Parnell, W. J. and Abrahams, I. D.},
  journal={SIAM J. Appl. Math.},
  volume={79},
  number={6},
  pages={2569--2592},
  year={2019},
  publisher={SIAM}
}

@article{gower2021effective,
doi = {10.1088/1367-2630/abdfee},
year = {2021},
publisher = {IOP Publishing},
volume = {23},
number = {6},
pages = {063083},
author = {Artur L Gower and Gerhard Kristensson},
title = {Effective waves for random three-dimensional particulate materials},
journal = {New Journal of Physics}
}

@inbook{Slab_book,
    author = {Gerhard Kristensson  and  Niklas Wellander},
    title = {Multiple scattering by a collection of randomly located obstacles distributed in a dielectric slab},
    publisher = {Advances in Mathematical Methods for Electromagnetics},
    year = {2020},
    chapter = {25},
    doi = {10.1049/SBEW528E_ch25}
}

@Book{Kristensson2016,
  author    = {Kristensson, Gerhard},
  title     = {Scattering of Electromagnetic Waves by Obstacles},
  year      = {2016},
  series    = {{Mario Boella Series on Electromagnetism in Information and Communication}},
  publisher = {SciTech Publishing},
  address   = {Edison, NJ, USA}
}

@InCollection{Bostrom+Kristensson+Strom1991,
  Title                    = {Transformation Properties of Plane, Spherical and Cylindrical Scalar and Vector Wave Functions},
  Author                   = {Bostr{\"o}m, Anders and Kristensson, Gerhard and Str{\"o}m, Staffan},
  Booktitle                = {Field Representations and Introduction to Scattering},
  Year                     = {1991},

  Chapter                  = {4},
  Editor                   = {Varadan, V. V. and Lakhtakia, A. and Varadan, V. K.},
  Pages                    = {165--210},
  Series                   = {Acoustic, Electromagnetic and Elastic Wave Scattering}
}

@article{10.1063/1.1704333,
    author = {Danos, M. and Maximon, L. C.},
    title = "{Multipole Matrix Elements of the Translation Operator}",
    journal = {Journal of Mathematical Physics},
    volume = {6},
    number = {5},
    pages = {766-778},
    year = {1965},
    issn = {0022-2488},
    doi = {10.1063/1.1704333}
}

@inbook{gerhard_thesis,
title = "Electromagnetic scattering from a buried three-dimensional inhomogeneity in a lossy ground",
author = "Gerhard Kristensson",
year = "1979",
language = "English",
series = "Technical Report, Institute of Theoterical Physics, Chalmers University of Technology",
number = "79-29",
booktitle = "Technical Report",
}

@article{Caleap_2012,
doi = {10.1088/1367-2630/14/3/033014},
year = {2012},
publisher = {IOP Publishing},
volume = {14},
number = {3},
pages = {033014},
author = {Mihai Caleap and Bruce W Drinkwater and Paul D Wilcox},
title = {Effective dynamic constitutive parameters of acoustic metamaterials with random microstructure},
journal = {New Journal of Physics}
}

@article{Caleap_2015,
doi = {10.1088/1367-2630/17/12/123022},
url = {https://dx.doi.org/10.1088/1367-2630/17/12/123022},
year = {2015},
publisher = {IOP Publishing},
volume = {17},
number = {12},
pages = {123022},
author = {Mihai Caleap and Bruce W Drinkwater},
title = {Metamaterials: supra-classical dynamic homogenization*},
journal = {New Journal of Physics}
}

@article{Linton_2006,
author = {Linton, C. and Martin, Paul},
year = {2006},
pages = {},
title = {Multiple Scattering by Multiple Spheres: A New Proof of the Lloyd--Berry Formula for the Effective Wavenumber},
volume = {66},
journal = {SIAM Journal on Applied Mathematics},
doi = {10.1137/050636401}
}

@article{Linton_2005,
    author = {Linton, C. M. and Martin, P. A.},
    title = "{Multiple scattering by random configurations of circular cylinders: Second-order corrections for the effective wavenumber}",
    journal = {The Journal of the Acoustical Society of America},
    volume = {117},
    number = {6},
    pages = {3413-3423},
    year = {2005},
    doi = {10.1121/1.1904270}
}

@article{Parnell_2010,
author = {Parnell, William and Abrahams, Ian},
year = {2010},
pages = {678-701},
title = {Multiple point scattering to determine the effective wavenumber and effective material properties of an inhomogeneous slab},
volume = {20},
journal = {Waves in Random and Complex Media},
doi = {10.1080/17455030.2010.510858}
}

@book{ISO_2017,
    author = {ISO 36BI
20998-3:2017},
    title = { Measurement and Characterization of Particles by Acoustic Methods vol 3},
    publisher = {London: British Standards Institution},
    year = {2017}
}

@article{Challis_2005,
doi = {10.1088/0034-4885/68/7/R01},
year = {2005},
publisher = {},
volume = {68},
number = {7},
pages = {1541},
author = {R E Challis and M J W Povey and M L Mather and A K Holmes},
title = {Ultrasound techniques for characterizing colloidal dispersions},
journal = {Reports on Progress in Physics}
}

@article{FORRESTER_2016,
title = {Characterisation of colloidal dispersions using ultrasound spectroscopy and multiple-scattering theory inclusive of shear-wave effects},
journal = {Chemical Engineering Research and Design},
volume = {114},
pages = {69-78},
year = {2016},
issn = {0263-8762},
doi = {10.1016/j.cherd.2016.08.008},
author = {D.M. Forrester and J. Huang and V.J. Pinfield}
}

@article{MISHCHENKO_2016,
title = {First-principles modeling of electromagnetic scattering by discrete and discretely heterogeneous random media},
journal = {Physics Reports},
volume = {632},
pages = {1-75},
year = {2016},
note = {First-principles modeling of electromagnetic scattering by discrete and discretely heterogeneous random media},
issn = {0370-1573},
doi = {https://doi.org/10.1016/j.physrep.2016.04.002},
author = {Michael I. Mishchenko and Janna M. Dlugach and Maxim A. Yurkin and Lei Bi and Brian Cairns and Li Liu and R. Lee Panetta and Larry D. Travis and Ping Yang and Nadezhda T. Zakharova}
}

@article{Foldy_1945,
  title = {The Multiple Scattering of Waves. I. General Theory of Isotropic Scattering by Randomly Distributed Scatterers},
  author = {Foldy, Leslie L.},
  journal = {Phys. Rev.},
  volume = {67},
  issue = {3-4},
  pages = {107--119},
  numpages = {0},
  year = {1945},
  publisher = {American Physical Society},
  doi = {10.1103/PhysRev.67.107}
}

@book{Huang_1963,
publisher = {Wiley,},
isbn = {9780521432245},
year = {1963},
title = {Statistical mechanics},
address = {New York:},
author = {Huang, Kerson},
}

@book{kolomietz2020mean,
  title={Mean Field Theory},
  author={Kolomietz, V.M. and Shlomo, S.},
  isbn={9789811211799},
  year={2020},
  publisher={World Scientific Publishing Company}
}

@incollection{2016_closure,
  author      = "Christian Kuehn",
  title       = "Moment Closure-A Brief Review",
  booktitle   = "Control of Self-Organizing Nonlinear Systems",
  publisher   = "Springer International Publishing",
  year        = 2016,
  pages       = "253-271",
  chapter     = 13,
  DOI={10.1007/978-3-319-28028-8}
}

@article{Gower_2023,
  title = {A model to validate effective waves in random particulate media: spherical symmetry},
  author = {Gower, Artur L. and Hawkins, Stuart C. and Kristensson, Gerhard},
  journal = {Proc. R. Soc. A},
  volume = {479:},
  issue = {20230444},
  year = {2023},
  doi = {10.1098/rspa.2023.0444}
}

@ARTICLE{Closure1971,
       author = {{Adomian}, G.},
        title = "{The closure approximation in the hierarchy equations}",
      journal = {Journal of Statistical Physics},
         year = 1971,
       volume = {3},
       number = {2},
        pages = {127-133},
          doi = {10.1007/BF01019846},
       adsurl = {https://ui.adsabs.harvard.edu/abs/1971JSP.....3..127A}
}

@article{Lax_QCA_1952,
  title = {Multiple Scattering of Waves. II. The Effective Field in Dense Systems},
  author = {Lax, Melvin},
  journal = {Phys. Rev.},
  volume = {85},
  issue = {4},
  pages = {621--629},
  numpages = {0},
  year = {1952},
  publisher = {American Physical Society},
}

@article{Martin_2020,
    author = {Martin, P. A. and Skvortsov, A. T.},
    title = "{Scattering by a sphere in a tube, and related problems}",
    journal = {The Journal of the Acoustical Society of America},
    volume = {148},
    number = {1},
    pages = {191-200},
    year = {2020},
    issn = {0001-4966},
    doi = {10.1121/10.0001518}
}

@article{Waterman1971,
  title = {Symmetry, Unitarity, and Geometry in Electromagnetic Scattering},
  author = {Waterman, P. C.},
  journal = {Phys. Rev. D},
  volume = {3},
  issue = {4},
  pages = {825--839},
  numpages = {0},
  year = {1971},
  publisher = {American Physical Society},
  doi = {10.1103/PhysRevD.3.825},
  url = {https://link.aps.org/doi/10.1103/PhysRevD.3.825}
}

@article{Varadan1978,
    author = {Varadan, Vijay K. and Varadan, Vasundara V. and Pao, Yih‐Hsing},
    title = "{Multiple scattering of elastic waves by cylinders of arbitrary cross section. I. SH waves}",
    journal = {The Journal of the Acoustical Society of America},
    volume = {63},
    number = {5},
    pages = {1310-1319},
    year = {1978},
    issn = {0001-4966},
    doi = {10.1121/1.381883}
}

@article{Martin2008,
author = {Martin, Paul and Maurel, Agnès},
year = {2008},
pages = {865-880},
title = {Multiple scattering by random configurations of circular cylinders: Weak scattering without closure assumptions},
volume = {45},
journal = {Wave Motion},
doi = {10.1016/j.wavemoti.2008.03.004}
}

@article{Aris2024,
doi = {10.1088/1367-2630/ad49c2},
year = {2024},
publisher = {IOP Publishing},
volume = {26},
number = {6},
pages = {063002},
author = {Aristeidis Karnezis and Paulo S Piva and Art L Gower},
title = {The average transmitted wave in random particulate materials},
journal = {New Journal of Physics}
}

@article{Kevish2024,
author = {Napal, K. K.  and Piva, P. S.  and Gower, A. L. },
title = {Effective T-matrix of a cylinder filled with a random two-dimensional particulate},
journal = {Proceedings of the Royal Society A: Mathematical, Physical and Engineering Sciences},
volume = {480},
number = {2292},
pages = {20230660},
year = {2024},
doi = {10.1098/rspa.2023.0660}
}

@article{Varadans_84,
title = {Multiple scattering theory for wave propagation in discrete random media},
journal = {International Journal of Engineering Science},
volume = {22},
number = {8},
pages = {1139-1148},
year = {1984},
issn = {0020-7225},
doi = {10.1016/0020-7225(84)90115-0},
author = {Y. Ma and V.V. Varadan and V.K. Varadan}
}

@article{Simon_tony2024,
    author = {Simon, Alverède and Baudis, Quentin and Wunenburger, Régis and Valier-Brasier, Tony},
    title = "{Propagation of elastic waves in correlated dispersions of resonant scatterers}",
    journal = {The Journal of the Acoustical Society of America},
    volume = {155},
    number = {6},
    pages = {3627-3638},
    year = {2024},
    doi = {10.1121/10.0026233}
}

@article{Hynne_1987,
author = {Hynne, F.  and Bullough, R. K.  and Edwards, Samuel Frederick },
title = {The scattering of light. II. The complex refractive index of a molecular fluid},
journal = {Philosophical Transactions of the Royal Society of London. Series A, Mathematical and Physical Sciences},
volume = {321},
number = {1559},
pages = {305-360},
year = {1987},
doi = {10.1098/rsta.1987.0017}
}

@article{Mishchenko_2018,
title = {Overview of methods for deriving the radiative transfer theory from the Maxwell equations. I: Approach based on the far-field Foldy equations},
journal = {Journal of Quantitative Spectroscopy and Radiative Transfer},
volume = {220},
pages = {123-139},
year = {2018},
issn = {0022-4073},
doi = {https://doi.org/10.1016/j.jqsrt.2018.09.004},
author = {Adrian Doicu and Michael I. Mishchenko}
}

@article{Mishchenko_2011,
title = {Scattering of electromagnetic waves by ensembles of particles and discrete random media},
journal = {Journal of Quantitative Spectroscopy and Radiative Transfer},
volume = {112},
number = {13},
pages = {2095-2127},
year = {2011},
note = {Polarimetric Detection, Characterization, and Remote Sensing},
issn = {0022-4073},
doi = {10.1016/j.jqsrt.2011.04.010},
author = {Victor P. Tishkovets and Elena V. Petrova and Michael I. Mishchenko}
}

@article{Al-Lashi_2015,
    author = {Al-Lashi, Raied S. and Challis, Richard E.},
    title = "{Ultrasonic particle sizing in aqueous suspensions of solid particles of unknown density}",
    journal = {The Journal of the Acoustical Society of America},
    volume = {138},
    number = {2},
    pages = {1023-1029},
    year = {2015},
    issn = {0001-4966},
    doi = {10.1121/1.4927694}
}

@misc{Miyamoto_1999,
title = {Functionally graded materials: Design, processing and applications},
author = {Miyamoto, Y and Kaysser, W A and Rabin, B H and Kawasaki, A and Ford, R G},
place = {United States},
year = {1999}
}

@article{EffectiveWaves.jl,
  title={Effectivewaves.jl: A julia package to calculate ensemble averaged waves in heterogeneous materials.},
  author={Artur L. Gower},
  journal={ \href{https://github.com/JuliaWaveScattering/EffectiveWaves.jl}{Github}},
  year={2020}
}

@inproceedings{twersky1964,
  title={On propagation in random media of discrete scatterers},
  author={Twersky, Victor},
  booktitle={Proc. Symp. Appl. Math},
  volume={16},
  pages={84--116},
  year={1964}
}

@article{Garcia-Valenzuela:05,
author = {Augusto Garc\'{i}a-Valenzuela and Rub\'{e}n G. Barrera and Celia S\'{a}nchez-P\'{e}rez and Alejandro Reyes-Coronado and Eugenio R. M\'{e}ndez},
journal = {Opt. Express},
number = {18},
pages = {6723--6737},
publisher = {Optica Publishing Group},
title = {Coherent reflection of light from a turbid suspension of particles in an internal-reflection configuration: Theory versus experiment},
volume = {13},
year = {2005},
doi = {10.1364/OPEX.13.006723}
}

@article{Chung_2011,
    author = {Chung, Chul-Woo and Popovics, John S. and Struble, Leslie J.},
    title = "{Flocculation and sedimentation in suspensions using ultrasonic wave reflection}",
    journal = {The Journal of the Acoustical Society of America},
    volume = {129},
    number = {5},
    pages = {2944-2951},
    year = {2011},
    issn = {0001-4966},
    doi = {10.1121/1.3569730}
}

@article{Li_2018,
doi = {10.1088/1757-899X/394/2/022065},
year = {2018},
publisher = {IOP Publishing},
volume = {394},
number = {2},
pages = {022065},
author = {Weikai Li and Baohong Han},
title = {Research and Application of Functionally Gradient Materials},
journal = {IOP Conference Series: Materials Science and Engineering}
}

\end{document}